\documentclass[11pt]{article}
\pdfoutput=1
\usepackage{jheppub} 
\usepackage{lineno}
\DeclareMathOperator{\Tr}{Tr}

\usepackage{float}
\usepackage{braket}
\usepackage{mathtools}
\usepackage{fancyhdr}
\usepackage{mathrsfs}
\usepackage{physics}
\usepackage{tikz}
\usepackage{mathdots}
\usepackage{yhmath}
\usepackage{makecell}
\usepackage{cancel}
\usepackage{circuitikz}
\usepackage{color}
\usepackage{tikz-cd}
\usepackage{array}
\usepackage{placeins}
\usepackage{multirow}
\usepackage{tabularx}
\usepackage{extarrows}
\usepackage{booktabs}
\usetikzlibrary{fadings}
\usetikzlibrary{patterns}
\usetikzlibrary{shadows.blur}
\usetikzlibrary{shapes}
\usepackage{graphicx}
 \usepackage{verbatim} 
\newcommand{\llangle}{\langle\!\langle}
\newcommand{\rrangle}{\rangle\!\rangle}

\title{\boldmath Symmetry Resolved Entanglement Entropy: Equipartition under Driven and Non-unitary Evolution in a Compact Boson CFT}

\author[a]{Filiberto Ares,}
\author[b]{Jayashish Das,}
\author[b]{Arnab Kundu}

\affiliation[a]{SISSA and INFN Sezione di Trieste, via Bonomea 265, 34136 Trieste, Italy}
\affiliation[b]{Theory Division, Saha Institute of Nuclear Physics, A CI of Homi Bhabha National Institute, 1/AF, Bidhannagar, Kolkata 700064, India}
\emailAdd{faresase[at]sissa.it}
\emailAdd{jayashish.das[at]saha.ac.in}
\emailAdd{arnab.kundu[at]saha.ac.in}

\abstract{We study the evolution of symmetry-resolved entanglement entropy in bulk-driven Floquet conformal field theories (CFTs). Focusing on the two-dimensional free compact boson CFT, we analyze how symmetry-resolved Rényi entropies approach or depart from equipartition among charge sectors. We show that the existence of an $\mathfrak{sl}^{(k)}(2,\mathbb{R})$ subalgebra of the Virasoro algebra introduces a free parameter, the label $k$, which allows us to control the breakdown of equipartition. We argue that this effect originates from an explicit coupling between low- and high-frequency modes. Based on a general oscillator representation of the Virasoro algebra, we expect this mechanism to persist beyond the free boson CFT. Finally, we discuss how the real-time dynamics of fine-grained symmetry-resolved entropies of a boundary state are modified under non-unitary evolution, which can be associated with post-selected weak measurements.}

\begin{document}
\maketitle
\flushbottom

\section{Introduction}

Entanglement is fundamental for understanding the properties of extended quantum systems. One key feature is that it encodes universal data of the conformal field theories describing critical points. These data can be extracted from different entanglement measures, with R\'enyi entropies playing a prominent role. A central result in 1+1-dimensional CFT is that the entanglement entropy of an interval in the ground state grows logarithmically with the interval size, with a prefactor proportional to the central charge of the theory~\cite{Holzhey94, cc-04, cc-09}. Furthermore, the entanglement of several disjoint intervals is determined by the partition function of the CFT on higher-genus surfaces and therefore depends on its field content~\cite{Caraglio08, Furukawa08, calabrese09disjoint, calabrese11disjoint}. Entanglement entropies also provide valuable insights into out-of-equilibrium systems. In particular, they serve as a probe of the relaxation dynamics of isolated many-body quantum systems following quantum quenches~\cite{calabrese05evolution, kaufman16thermalization, alba17thermo}. Remarkably, several crucial features observed in such quenches were first identified in CFTs~\cite{Calabrese_2006, calabrese07quenches, calabrese07local, Calabrese:2016xau}. Periodically driven many-body systems, or Floquet dynamics, provide another key out-of-equilibrium setting where CFT methods have proven useful. Floquet systems have attracted considerable attention because they can give rise to non-equilibrium phenomena with no equilibrium counterpart, including Floquet topological insulators~\cite{Oka2019-ep,Lindner:2011guf}, protected phases~\cite{PhysRevX.6.041001,NathanRudner2015}, and Floquet time crystals~\cite{khemani2019briefhistorytimecrystals,Else_2020,PhysRevLett.117.090402,Zaletel_2023,PhysRevLett.118.030401}. 

In this context, periodically bulk-driven CFTs represent an analytically solvable setup to understand Floquet dynamics, since the evolution of correlation functions and entanglement entropy can be analytically determined~\cite{wen2018floquetconformalfieldtheory, Fan_2020, PhysRevResearch.2.023085, Han:2020kwp, PhysRevB.103.224303, Wen:2020wee, PhysRevResearch.2.033461, Choo:2022lgm, Fan_2021, Wen:2021mlv, Wen:2022pyj, Das:2022jrr, Das:2022pez,  Das_2024, Lapierre25engineered, Das:2025wjo,Erdmenger:2025chu,Fang:2025rie}. In particular, the time evolution protocol alternates between Hamiltonians belonging to a family of spatially inhomogeneous deformed CFTs. Each Hamiltonian is associated with an $\mathfrak{sl}^{(k)}(2, \mathbb{R})$ subalgebra of the full Virasoro algebra of the homogeneous CFT. The entanglement entropy reveals two phases depending on the driving frequency: a heating phase, characterized by linear growth in time, and a non-heating phase, where the entropy oscillates. The phase transition separating them is characterized by logarithmic growth in time. These phases also correspond one-to-one to the conjugacy classes of the Möbius group governing the evolution of the primary fields in each Floquet cycle.

An interesting variant in both quantum quenches and driven protocols can be realized when the time evolution becomes non-unitary as a result of interactions with the environment or measurements. There are several ways to implement non-unitary dynamics, for instance through Lindblad master equations or non-Hermitian Hamiltonians~\cite{mo25complex}. Another recent approach in quantum field theory is to consider a complex Lorentzian time~\cite{wen2024exactly}. This idea is formally motivated by the Kontsevich–Segal–Witten (KSW) consistency criteria for quantum field theories defined on complex spacetime geometries, which allow for this type of time evolution~\cite{kontsevich21wick, Witten:2021nzp}. From a physical perspective, imaginary-time evolution can be understood as a post-selected weak measurement of the system’s energy, which preserves conformal invariance. At equilibrium, weak measurements can substantially reshape the properties of critical ground states~\cite{garratt23measurements, yang23measurements, murciano23measurements, weinstein23measurements, sala24measurement, patil24measurements, liu25measurements}. Out of equilibrium, alternating real and imaginary time evolution gives rise to a competition between the two, which may lead to distinct steady-state, measurement-induced phases, and other phenomena~\cite{su24dynamics, Mao24Local, Lapierre:2025zsg, Bai:2026avl}.

Recently, our understanding of entanglement has greatly benefited from accounting for the global symmetries of the theory. The decomposition of the entanglement entropy into charge sectors gives a more complete view of the structure of entanglement in many-body quantum states~\cite{laflorencie14spin, PhysRevLett.120.200602, xavier2018, lukin19exp}. In many cases, the entanglement is at leading order uniformly distributed over symmetry sectors, a phenomenon known as entanglement equipartition~\cite{xavier2018}. The main observable to study the interplay between symmetries in entanglement is the symmetry-resolved entanglement entropy, which has been studied in multiple settings, including integrable spin chains~\cite{bonsignori19, Fraenkel19sym, Murciano:2019wdl,  Bonsignori:2020laa}, QFTs~\cite{Murciano:2020vgh, Horvath20boost, Horvath21U1, Capizzi:2022jpx,  Castro-Alvaredo:2024azg}, and CFTs~\cite{capizzi20excited, Hung21branes, Calabrese_2021, estienne21corrections, Ares:2022gjb, Milekhin23charge, Ghasemi23universal, DiGiulio:2022jjd, Kusuki:2023bsp, foligno23torus, capizzi23defect, PhysRevLett.131.151601, fossati23nonherm, Gaur:2023yru, Gaur:2024vdh, Northe:2025qcv,Bai:2025ysg,Mao:2025hkp}, not only at equilibrium but also following quantum quenches~\cite{parez21quasiparticle, Parez:2021pgq, Feldman:2019upn, scopa22hydro, Li25ballistic, Chen:2025fzm, Zhao:2024ivy,Li:2024qpx}. 

Here we study the evolution of the symmetry-resolved entanglement entropy in bulk-driven Floquet CFTs. As a concrete example, we consider the compact boson CFT, which possesses a ${\rm U(1)}$ global symmetry. We find that when the theory is driven by spatially inhomogeneous deformed Hamiltonians, the label $k$ of the $\mathfrak{sl}^{(k)}(2,\mathbb{R})$ Virasoro subalgebra associated with them sets the scale at which equipartition of entanglement among the charge sectors is satisfied. We also analyze the evolution of the symmetry-resolved entanglement entropy following a quench with complex Lorentzian time in an infinite segment. We find that, although their long-time behavior respects equipartition at leading order, it differs from that of the unitary case~\cite{parez21quasiparticle}.

The paper is organized as follows. In Sec.~\ref{sec:sree}, we define the symmetry-resolved entanglement entropies and review the main tools used to calculate them. In Sec.~\ref{sec:compact_boson}, we revisit some basic aspects of the compact boson CFT that will be useful for interpreting our results. In Sec.~\ref{eq:driven_cft}, we first introduce the class of spatially inhomogeneous Hamiltonians used to drive the system and discuss some basic results of driven CFTs. We then analyze the symmetry-resolved entanglement entropies when the system is driven by these Hamiltonians. In Sec.~\ref{sec:nonunitary_quench}, we examine the evolution of the symmetry-resolved entropies after a complex-time quench. Finally, in Sec.~\ref{sec:sum_discuss}, we draw conclusions and discuss future prospects.

\section{Symmetry Resolved Entanglement Entropy}\label{sec:sree}

Let us consider an extended quantum system in a pure state $|\Psi\rangle$, that can be divided into two spatial parts $A\cup B$, such that the total Hilbert space factorises as: ${\mathcal H}= {\mathcal H}_A \otimes {\mathcal H}_B$. The
state of subsystem $A$ is described by the reduced density matrix: $\rho_A = {\rm Tr}_B\left( | \Psi\rangle \langle\Psi | \right)$, where ${\rm Tr}_B$ is the partial trace over $B$. The R\'{e}nyi-$n$ entanglement entropy is defined as:
\begin{eqnarray}
    S_A^{(n)} = \frac{1}{1-n} \log {\rm Tr} \left( \rho_A^n\right) \ .\label{eq:sree-renyi-entropy}
\end{eqnarray}
The limit $n\to 1$ gives the (von Neumann) entanglement entropy (EE):
\begin{eqnarray}
    S_A^{(1)} = - {\rm Tr} \left( \rho_A \log \rho_A \right) \ . \label{eq:sree-von-neumann-entropy}
\end{eqnarray}

Let us now further assume that the system has a global U(1) symmetry generated by a
charge, $Q$, which can be decomposed into the sum of the charges in $A$ and in $B$. That is, $Q = Q_A + Q_B$. If $|\Psi\rangle$ is an eigenstate of $Q$, then the density matrix $\rho \equiv | \Psi\rangle \langle\Psi |$ commutes with the charge operator: $[Q, \rho]=0$. Taking the partial trace over the $B$ subsystem, we obtain:
\begin{eqnarray}
    {\rm Tr}_B[Q, \rho]=0 \quad \implies \quad [Q_A, \rho_A]=0 \ .\label{eq:sree-qa-commutes-rhoa}
\end{eqnarray}
This property implies that $\rho_A$ is endowed with a block diagonal structure, in which each block
corresponds to an eigenvalue, denoted by $q$, of the operator $Q_A$; that is:
\begin{eqnarray}
    \rho_A = \bigoplus_{q} p_q \rho_{A, q} \ ,\label{eq:sree-rhoa-block-decomposition}
\end{eqnarray}
where $p_q$ is the probability of finding a charge $q$ in a measurement of $Q_A$. Suppose $\Pi_q$ is the
projector onto the eigenspace associated with the eigenvalue $q$, then $p_q = {\rm Tr}(\Pi_{q} \rho_A)$. The density matrices
\begin{eqnarray}
    \rho_{A, q} \equiv \frac{\Pi_q \rho_A \Pi_q }{p_q} \ ,\label{eq:sree-rhoa-q-definition}
\end{eqnarray}
describe the state of $A$ in the charge sector $q$ and are normalised in the
sense that: ${\rm Tr}(\rho_{A,q}) = 1$.

The amount of entanglement between $A$ and $B$ in each charge sector can be quantified
by the symmetry-resolved R\'{e}nyi entropies~\cite{PhysRevLett.120.200602, xavier2018}:
\begin{eqnarray}
    S_{A,q}^{(n)} = \frac{1}{1-n} \log {\rm Tr} \left( \rho_{A,q}^n\right) \ .\label{eq:sree-symmetry-resolved-renyi}
\end{eqnarray}
Subsequently, taking the limit $n \to 1$, we obtain the symmetry-resolved entanglement entropy (SREE):
\begin{eqnarray}
    S_{A,q}^{(1)} = - {\rm Tr} \left( \rho_{A,q} \log \rho_{A,q} \right) \ .\label{eq:sree-symmetry-resolved-vn}
\end{eqnarray}
Using the decomposition in (\ref{eq:sree-rhoa-block-decomposition}), the total entanglement entropy of (\ref{eq:sree-von-neumann-entropy}) is given by~\cite{lukin19exp, xavier2018}
\begin{eqnarray}
    S_A^{(1)} = \sum_q p_q S_{A,q}^{(1)} - \sum_q p_q \log p_q \equiv S_A^{\rm c} + S_A^{\rm num} \ .\label{eq:sree-entropy-decomposition-config-number}
\end{eqnarray}
Here $S_A^{\rm c}$ is the configurational entropy — it quantifies the average contribution to the total entanglement of all the charge sectors — and $S_A^{\rm num}$ is the number entropy — it takes
into account the entanglement due to the fluctuations of the value of the charge within $A$.

For explicit calculations, we begin with the following representation of the projection operator:
\begin{eqnarray}
    \Pi_q =\int_{-\pi}^{\pi} \frac{d\alpha}{2\pi} e^{i \alpha(Q_A-q)}  \ .\label{eq:sree-projector-fourier}
\end{eqnarray}
Using this, we can re-express the symmetry-resolved R\'{e}nyi entropy as~\cite{PhysRevLett.120.200602, xavier2018}:
\begin{eqnarray}
    S_{A,q}^{(n)} = \frac{1}{1-n} \log \left[ \frac{\mathcal Z_n(q)}{\mathcal Z_1(q)^n}\right] \ ,\label{eq:sree-renyi-via-Znq}
\end{eqnarray}
where
\begin{eqnarray}
    \mathcal Z_n(q) \equiv {\rm Tr} \left( \Pi_q \rho_A^n\right)  = \int_{-\pi}^\pi \frac{d\alpha}{2\pi} e^{- i \alpha q} Z_n(\alpha) \ , \quad Z_n(\alpha) = {\rm Tr}\left(\rho_A^n e^{i \alpha Q_A} \right) \ .\label{eq:sree-Znq-and-Znalpha-definitions}
\end{eqnarray}
and $Z_n(\alpha)$ are the charged moments of the reduced density matrix $\rho_A$.

Now we can write the symmetry-resolved von Neumann Entropy by taking the limit $n \rightarrow1$ in \eqref{eq:sree-renyi-via-Znq} ,
\begin{equation}
    S_{A,q}=\log{\mathcal Z_1(q)}-\cfrac{1}{\mathcal Z_1(q)}\partial_n\mathcal Z_n(q)\Bigg|_{n=1} \ .\label{eq:sree-vn-via-Znq}
\end{equation}

Using the path integral representation of $\rho_A$, the moments at vanishing charge, $Z_n(0) = {\rm Tr}\left(\rho_A^n \right)$, are the partition functions of the CFT on an $n$-sheeted Riemann surface with branch cuts, in which the sheets are sewn together along the interval $A\equiv [u,v]$ in a cyclic manner~\cite{cc-04, cc-09}. As a result, the partition function $Z_n(0)$ can be expressed as the two-point correlation function on the complex plane of the twist and anti-twist fields $\tau_n$ and $\tilde{\tau}_n$, which are inserted at the end-points
of $A$:
\begin{eqnarray}
  Z_n(0) \equiv \left \langle\tau_n(u) \tilde{\tau}_n(v)  \right \rangle \  \quad {\rm with} \quad h_n = \bar{h}_n= \frac{c}{24} \left( n - \frac{1}{n}\right) \ .\label{eq:sree-twist-correlator-neutral}
\end{eqnarray}

The charged moments, on the other hand, are dressed with a flux line that is threading through the $n$-sheeted Riemann surface. As a result, a charged particle that traverses a closed loop crossing all the sheets, acquires an Aharonov-Bohm phase $e^{i \alpha}$. This phase can be realized by inserting a local U$(1)$ vertex operator, denoted by $V_\alpha$, at the endpoints of the interval $A$. Therefore, the charged moments can now be realized in terms of composite twist fields, $\tau_{n,\alpha} \equiv \tau_n V_\alpha$,  which yields~\cite{PhysRevLett.120.200602}:
\begin{eqnarray}
    Z_n(\alpha) \equiv \left \langle\tau_{n,\alpha}(u) \tilde{\tau}_{n, -\alpha}(v)  \right \rangle  \quad {\rm with} \quad h_{n,\alpha} = \frac{c}{24} \left( n - \frac{1}{n}\right) + \frac{h_{\alpha}^V}{n} \ . \label{eq:sree-twist-correlator-charged}
\end{eqnarray}
The shift in the dimension for the composite twist operator depends on the particular CFT. For example, consider a free compact boson $\varphi(x)$ defined on a circle of radius $R$, \textit{i.e.}, $\varphi \sim \varphi + 2 \pi R$. This theory has global U$(1)$ symmetry associated with the conserved current $j(x) = \partial\varphi(x)$. The corresponding charge in the interval $A$ reads
\begin{equation}
     Q_A \equiv \int_u^v dx \partial \varphi(x) = \varphi(u) - \varphi(v) \ .
\end{equation}
Thus, the insertion $e^{i\alpha Q_A}$ can be represented in terms of the pair of vertex operators
\begin{equation}
     e^{i \alpha Q_A} \equiv V_\alpha(u) V_{-\alpha}(v) \ , \quad V_\alpha(x) = e^{i\alpha \varphi(x)}  \label{eq:free-boson-charge-vertex} \ . 
\end{equation}
In this particular case, the shift~\eqref{eq:sree-twist-correlator-charged} in the dimension of the composite twist field is given by  $h_\alpha^V = \frac{\alpha^2}{\sqrt{2R}}$. In this article, we will primarily consider the free compact boson CFT.

\section{Some Basics of the Free Compact Boson}\label{sec:compact_boson}


Before moving further, let us discuss some basic aspects of the free $2d$ boson theory. Let us define the free boson theory on a cylinder, ${\mathbb R}\times S^1$, with explicit Euclidean coordinates $\{\tau, \sigma\}$. Let us choose $\tau \in {\mathbb R}$ and $\sigma \in S^1$, of circumference $L=2\pi$.\footnote{Henceforth, any dimensionful quantity will be measured in units of $L$, wherever explicit factors of $L$ do not appear.} Since this is the Euclidean cylinder, we can interchangeably consider $\tau$ and $\sigma$ as the corresponding Euclidean time.\footnote{When $\tau$ is the Euclidean time, we consider a free boson system on a finite spatial circle and when $\sigma$ is the Euclidean time, it corresponds to the thermal state with an inverse temperature $\beta = i L$.}

Let us consider the compact boson: $\varphi(\tau,\sigma) = \varphi(\tau,\sigma) + 2\pi R$, where $R$ is the compactification radius.\footnote{The standard terminology is that $\{\tau,\sigma\}$ are the worldsheet coordinates and $\varphi(\tau,\sigma)$ the target space coordinate.} On the cylinder, the target space compactness implies that: $\varphi(\tau, \sigma+ L) = \varphi(\tau, \sigma) + 2 \pi R \omega $, where $\omega \in {\mathbb Z}$ denotes the winding modes. The corresponding Hamiltonian of the free boson theory is simply given by
\begin{eqnarray}
    H = \frac{1}{4\pi} \int_0^{2\pi} d\sigma \left[ \left( \partial_\tau\varphi\right)^2 + \left( \partial_\sigma\varphi\right)^2 \right] \label{eq:boson-cylinder-hamiltonian} \ .
\end{eqnarray}

The explicit mode-expansion, obtained from solving the resulting equations of motion, takes the standard form:
\begin{eqnarray}
    \varphi(\tau, \sigma) = \varphi_0 + \frac{n}{R}\tau + \omega R \sigma + i \sum_{k\not=0} \left[\alpha_k e^{- i k (\tau + \sigma) } + \tilde{\alpha}_k e^{-ik(\tau-\sigma)} \right] \label{eq:boson-cylinder-mode-expansion} \ ,
\end{eqnarray}
where $(n/R)$ corresponds to the quantized zero modes and $\{\alpha_k, \tilde{\alpha}_k\}$ are the oscillator modes. With the following identifications of the left-moving and the right-moving momenta and the corresponding Virasoro generators:
\begin{eqnarray}
  &&  p_L = \frac{n}{R} + \omega R \ , \quad p_R = \frac{n}{R} - \omega R \ , \\ \label{eq:boson-left-right-momenta}
  &&  L_0 = \frac{1}{2}p_L^2 + \sum_{k>0} \alpha_{-k} \alpha_k \ , \quad \bar{L}_0 = \frac{1}{2}p_R^2 + \sum_{k>0} \tilde{\alpha}_{-k} \tilde{\alpha}_k \ , \label{eq:boson-L0-L0bar-modes}
\end{eqnarray}
we obtain the standard CFT Hamiltonian on the cylinder:
\begin{eqnarray}
    H = L_0 + {\bar L}_0 - \frac{c}{12} \ ,\label{eq:boson-cylinder-hamiltonian-casimir}
\end{eqnarray}
where the $(-1/12)$ term comes from the Casimir energy on the cylinder.

Let us now consider a more general linear combination of the Virasoro generators as the Hamiltonian, \textit{i.e.} $H=L_k+L_{-k}+{\rm h.c.}$. Towards this, recall that, on the plane:
\begin{eqnarray}
    L_n = \frac{1}{2\pi i} \oint dz z^{n+1} T(z) \ , \quad T(z ) = - \frac{1}{2} \left( \partial\varphi(z) \right)^2 \ ,\label{eq:boson-virasoro-contour-stress-tensor}
\end{eqnarray}
and similarly for the anti-holomorphic coordinates. Using the plane to cylinder map: $z = e^{\tau + i \sigma}$ and $\bar{z}= e^{\tau - i \sigma}$, we obtain:
\begin{eqnarray}
    L_n = \frac{1}{2}\int_0^{2\pi} \frac{d\sigma}{2\pi} e^{i n \sigma} \left(\partial_+\varphi \right)^2    \ , \quad  \bar {L}_n = \frac{1}{2}\int_0^{2\pi} \frac{d\sigma}{2\pi} e^{-i n \sigma} \left(\partial_-\varphi \right)^2 \ , \quad \partial_{\pm} \equiv \partial_\tau \pm \partial_\sigma \ .\label{eq:boson-virasoro-modes-on-cylinder}
\end{eqnarray}
With these, it is now straightforward to check that:
\begin{eqnarray}\label{eq:Ham_Lk}
    && H = L_k + L_{-k} + {\rm h.c.} \nonumber\\
    && \implies \quad H = \frac{1}{2}\int_0^{2\pi} \frac{d\sigma}{2\pi} \cos(k\sigma) \left[\left( \partial_\tau \varphi\right)^2 +  \left( \partial_\sigma \varphi\right)^2 \right] \ .
\end{eqnarray}
It is straightforward to see that, because of the $\cos(k\sigma)$ factor in front, the Hamiltonian cannot be bounded from below or from above. Unlike the case of $k=0$, the oscillator basis is no longer the diagonal basis of the Hamiltonian, and schematically, it takes the form:
\begin{eqnarray}\label{eq:ham_osc_mod}
H \sim \sum_{m} \left(  \alpha_{k-m} \alpha_m + \alpha_{-k-m} \alpha_m\right)  +{\rm h.c.} \ ,
\end{eqnarray}
which clearly indicates how the $\alpha_m$ modes couple to the $\alpha_{k-m}$ modes, {\it etc}. 

According to Eq.~\eqref{eq:ham_osc_mod}, there is a non-trivial mixing between the low-lying IR modes with the UV-modes. Consider, for example,  $m=0$. The Hamiltonian contains an explicit coupling between modes $\alpha_k \alpha_0 $ and $\alpha_{-k} \alpha_0$. Thus, taking $k\gg 1$ demonstrates an explicit coupling of otherwise widely separated modes in the Fourier space. In the position space, {\it i.e.}~in terms of the worldsheet $\sigma$ coordinate, the same information is captured by how fast the cosine function in Eq.~\eqref{eq:Ham_Lk} oscillates.

The argument above generalizes purely in terms of the Virasoro algebra. This generalization is perhaps most simply seen by considering the oscillator representation of the Virasoro algebra, which was originally introduced in Ref.~\cite{Zamolodchikov:1986}, see also Refs.~\cite{Be_ken_2020,Besken:2019jyw} for more recent and explicit applications of the oscillator formalism. Given the free boson mode expansion:
\begin{eqnarray}
    \varphi = i \sum_{m=-\infty}^{\infty} \alpha_m z^{-m-1} \ , \quad {\rm with} \quad [\alpha_m, \alpha_n] = m \delta_{m,-n} \ , \label{eq:boson-oscillator-expansion-and-commutator}
\end{eqnarray}
the Virasoro generators are obtained from the stress-tensor of the linear dilaton theory:
\begin{eqnarray}
    T(z) = \frac{1}{2}\partial\varphi \partial\varphi + V \partial^2\varphi \equiv \sum_{m=-\infty}^\infty L_m z^{-m-2}\ . \label{eq:boson-linear-dilaton-stress-tensor-modes}
\end{eqnarray}
Explicitly written, one obtains, for example:
\begin{eqnarray}
  &&  L_k = \frac{1}{2}\sum_{n=-\infty}^\infty \alpha_{k - n } \alpha_n + i (k+1) V \alpha_k \ ,\label{eq:virasoro-oscillator-Lk} \\
  && L_0 = \frac{1}{2} \left(\alpha_0 \right)^2 + \sum_{k=1}^\infty \alpha_{-k} \alpha_k + i V \alpha_0 \label{eq:boson-virasoro-oscillator-L0}\ .
\end{eqnarray}
For $k\not=0$, we observe an explicit coupling between the $(k-m)$-th mode and the $m$-th mode, similar to above. The general oscillator representation of the Virasoro algebra is constructed using the above linear dilaton stress-tensor, upon introducing an infinitely many complex variables, denoted by $u_n$, such that $\alpha_n \sim \partial_{u_n}$ and $\alpha_{-n}\sim u_n$. Since the oscillator is just a property of the Virasoro algebra, we can run the logic backwards and conclude that the Lüscher-Mack type Hamiltonian will necessarily imply a coupling between the UV and the IR modes in a suitably written Hamiltonian. Therefore, we expect that the same effective length, associated with the hierarchy, will emerge in any CFT and our conclusions will also hold in general.

Before concluding this section, a few more comments are in order. First, note that, we can also define the compact boson theory on a strip. For example, we begin with the same Hamiltonian as before:
\begin{eqnarray}
    H = \frac{1}{4\pi} \int_0^L d\sigma \int d\tau \left[ \left( \partial_\tau \varphi\right)^2 + \left( \partial_\sigma \varphi\right)^2 \right] \ , \label{eq:boson-strip-hamiltonian}
\end{eqnarray}
with $\varphi = \varphi + 2 \pi R$. However, there are boundaries at $\sigma = 0 , L$ where we need to impose boundary conditions. We can use either a Dirichlet boundary condition, $\varphi=\varphi_0$ or a Neumann one: $\partial_\sigma \varphi = 0$ and obtain an explicit solution associated with these.\footnote{Of course, since there are two end-points, there are four inequivalent choices of boundary conditions: Dirichlet-Dirichlet, Neumann-Neumann, Dirichlet-Neumann and Neumann-Dirichlet. It is possible to write down the explicit mode expansions in each of these cases. However, we will not present them here, since these can be found in standard references. Nonetheless, note that, for Dirichlet-Dirichlet boundary conditions, there are no zero momentum modes but there are non-vanishing quantized winding modes. On the other hand, for Neumann-Neumann boundary condition, no winding modes are allowed while the momentum modes are quantized. Mixed boundary conditions remove all zero modes and render them half-integer valued.} The self-evident shift-symmetry of the Hamiltonian under $\varphi\to \varphi + \alpha$ becomes a U$(1)$-symmetry since $\alpha = \alpha + 2 \pi R$ for the compact boson. The conserved current $j^\mu \sim \partial^\mu \varphi$ now can have distinct behaviour at the boundaries, depending on the boundary conditions. For Dirichlet boundary conditions, $\partial^\sigma\varphi$ remains free at the boundary and therefore current conservation is violated. On the other hand, with the Neumann boundary condition at both ends, current conservation holds at the boundaries as well. Thus, on the strip, with the appropriate boundary conditions, the free boson is still endowed with a conserved U$(1)$-charge and correspondingly one can explore the symmetry resolved aspects of this sector, see Ref.~\cite{DiGiulio:2022jjd} for a detailed analysis.

\section{Symmetry-Resolved Entanglement Entropy in Driven CFT}\label{eq:driven_cft}

In this section, we study the evolution of the symmetry-resolved entanglement entropy~\eqref{eq:sree-symmetry-resolved-vn} in a driven $2d$ CFT. The system is initially prepared in the vacuum state $\ket{0}$ and subsequently evolved with a sequence of Hamiltonians $H_i$, each acting for a time interval $T_i$. The Hamiltonians $H_i$ belong to a specific family of inhomogeneous Hamiltonians, which will be described in detail below. After $N$ driving steps, the state of the system is
\begin{equation}\label{eq:driven_state}
\ket{\Psi_N} = U_N\ket{0}, \quad U_N=e^{-iT_N H_N}\cdots e^{-iT_2 H_2}e^{-iT_1 H_1}.
\end{equation}
Although one may consider different initial states, such as highly excited or thermal states, the main features of driven CFTs, including the structure of the phase diagram, are largely determined by the driving protocol rather than by the choice of the initial state~\cite{Wen:2020wee}. The evolution of symmetry-resolved entropies under a quench (i.e. when $N=1$ in Eq.~\eqref{eq:driven_state}) with the family of inhomogeneous Hamiltonians we are going to consider here has been recently studied in Ref.~\cite{Chen:2025fzm}

\subsection{Brief Review of Driven CFT}

We will begin by reviewing some basic aspects of driven $2d$ CFTs. Let us consider a $(1+1)d$ CFT defined on the cylinder ${\mathbb R}\times S^1$, whose time-evolution is governed by a family of inhomogeneous Hamiltonians\footnote{Sometimes, in the literature, these are referred to as the deformed Hamiltonians.} of the form~\cite{Wen:2020wee}:
\begin{equation}
H_i \;=\; \int_{0}^{L} \! \mathrm{d}x \; v_i(x)\,T_{00}(x) \ ,
\label{eq:drive-inhomogeneous-hamiltonian}
\end{equation}
where $L$ is the circumference of the cylinder, $T_{00}=(T(x)+\bar{T}(x))/(2\pi)$, and the deformation profile, denoted by $v_i(x)$, is chosen  as:
\begin{equation}
v_i(x)=a_i^{0}
+a_i^{+}\cos\!\left(\frac{2\pi x}{\ell}\right)
+a_i^{-}\sin\!\left(\frac{2\pi x}{\ell}\right) \ ,
\qquad a_i^{0},a_i^{+},a_i^{-}\in\mathbb{R} \ .
\label{eq:drive-deformation-profile-function}
\end{equation}
It is convenient to introduce the integer:
\begin{equation}
k \;=\; \frac{L}{\ell} \ ,
\qquad\text{so that}\qquad
\ell=\frac{L}{k} \ .
\label{eq:drive-k-and-ell-definition}
\end{equation}
With this choice, the deformation $v_i(x)$ couples only to Virasoro modes separated by $k$ units. In particular, if we split the Hamiltonian into the chiral and anti-chiral parts:
\begin{equation}
H_i \;=\; H_{i,\mathrm{chiral}}+H_{i,\mathrm{anti\text{-}chiral}} \label{eq:drive-chiral-antichiral-splitting}\ ,
\end{equation}
where the chiral contribution can be written as a general linear combination of $\{L_0,L_{\pm k}\}$ (and similarly $\{\bar L_0,\bar L_{\pm k}\}$ for the anti-chiral sector). A manifestly Hermitian parametrization uses the real combinations:\footnote{Here, the adjoint action is given by: $L_k^\dagger = L_{-k}$.}
\begin{equation}
L_{k,+}= \frac{1}{2}\big(L_k+L_{-k}\big) \ ,
\qquad
L_{k,-}=\frac{1}{2i}\big(L_k-L_{-k}\big) \ ,
\label{eq:drive-hermitian-virasoro-combinations}
\end{equation}
in terms of which we obtain:
\begin{equation}
H_{i,\mathrm{chiral}}
=\frac{2\pi}{L}\Big(a_i^{0}\,L_{0}+a_i^{+}\,L_{k,+}+a_i^{-}\,L_{k,-}\Big)
-\frac{\pi c}{12\,L} \ .
\label{eq:drive-chiral-hamiltonian-sl2k}
\end{equation}
Similarly, $H_{i,\mathrm{anti\text{-}chiral}}$ is obtained by replacing $L_k\to \bar L_k$.
The operators $\{L_0,L_{k,+},L_{k,-}\}$ form an $\mathfrak{sl}^{(k)}(2,\mathbb{R})$ algebra (embedded in the Virasoro algebra at level $k$) and, upon exponentiation, form a group which is often denoted as $\mathrm{SL}^{(k)}(2,\mathbb{R})$\footnote{Note that, locally, the $\mathfrak{sl}^{(k)}(2,\mathbb{R})$ algebra is isomorphic to the $\mathfrak{sl}(2,\mathbb{R})$ algebra, but, globally, they are not the same. Given the algebra:
\begin{eqnarray}
    \left[ L_k , L_0\right] = k L_k \ , \quad \left[ L_0 , L_{-k}\right] = k L_{-k} \ , \quad \left[ L_k , L_{-k}\right] = 2k L_0 + \frac{c}{12}k (k^2-1) \ ,
\end{eqnarray}
we can define:
\begin{eqnarray}
    {\cal L}_0 = \frac{1}{k}\left( L_0 + \frac{c}{24} (k^2 -1)\right)  \ , \quad {\cal L}_{\pm k} = \frac{1}{k} L_{\pm k} \ .
\end{eqnarray}
Here $\{{\cal L}_0, {\cal L}_{\pm k}\}$ satisfy an $\mathfrak{sl}(2,\mathbb{R})$ algebra~\cite{Jiang:2024hgt,Caputa:2022zsr,Anand:2017dav}.}.

A useful invariant that characterizes the $\mathfrak{sl}^{(k)}(2,\mathbb{R})$ element corresponding to the Hamiltonian~\eqref{eq:drive-chiral-hamiltonian-sl2k} is the quadratic form:
\begin{equation}
c^{(2)} \;=\; -\big(a^{0}\big)^2+\big(a^{+}\big)^2+\big(a^{-}\big)^2 \ ,
\label{eq:drive-sl2k-quadratic-invariant}
\end{equation}
(where we suppressed the step label $i$ for readability). The sign of $c^{(2)}$ determines the conjugacy class:
\begin{equation}
c^{(2)}<0:\ \text{elliptic} \ ,\qquad
c^{(2)}=0:\ \text{parabolic} \ ,\qquad
c^{(2)}>0:\ \text{hyperbolic} \ .
\label{eq:drive-conjugacy-classes-casimir-condition}
\end{equation}
In what follows, the square-root of this Casimir eigenvalue will become important. Keeping that in mind, let us define:
\begin{equation}
C \;\coloneqq\; \sqrt{\left|-\,(a_{0})^{2}+(a_{+})^{2}+(a_{-})^{2}\right|} \ .\label{eq:C-definition}
\end{equation}

The time evolution generated by $H_{\mathrm{chiral}}$ acts on the chiral coordinates through an $\mathrm{SU}(1,1)$ Möbius transformation\footnote{Note that, the $\mathrm{SU}(1,1)$ group is isomorphic to the $\mathrm{SL}(2,{\mathbb R})$ group. In this article, we will use them interchangeably.}. The corresponding evolution can therefore be represented by a matrix that depends on the time interval $t$:
\begin{equation}
M(t)=
\begin{pmatrix}
\alpha(t) & \beta(t)\\
\beta^*(t) & \alpha^*(t)
\end{pmatrix}\in \mathrm{SU}(1,1) \ ,
\qquad
|\alpha(t)|^2-|\beta(t)|^2=1 \ .
\label{eq:drive-su11-evolution-matrix}
\end{equation}
The three different conjugacy classes have three qualitatively distinct entries for these matrix elements, which take the following standard forms \cite{Wen:2022pyj}:
\begin{align}
\text{Elliptic:}\quad
&\left\{\begin{array}{l}\alpha(t) = -\cos\!\left(C\dfrac{\pi t}{\ell}\right)
             - i\,\dfrac{a_{0}}{C}\,\sin\!\left(C\dfrac{\pi t}{\ell}\right) \ ,\nonumber\\
\beta(t)  = -\,i\,\dfrac{a_{+}+i a_{-}}{C}\,\sin\!\left(C\dfrac{\pi t}{\ell}\right)\ , \end{array}\right. \nonumber\\
\text{Parabolic:}\quad
&\left\{\begin{array}{l}\alpha(t) = -1 - i\,a_{0}\,\dfrac{\pi t}{\ell},\\
 \beta(t)  = -\,i\,(a_{+}+i a_{-})\,\dfrac{\pi t}{\ell} \ , \end{array}\right.\nonumber\\
\text{Hyperbolic:}\quad
&\left\{\begin{array}{l}\alpha(t) = -\cosh\!\left(C\dfrac{\pi t}{\ell}\right)
             - i\,\dfrac{a_{0}}{C}\,\sinh\!\left(C\dfrac{\pi t}{\ell}\right) \ ,\nonumber\\
\beta(t)  = -\,i\,\dfrac{a_{+}+i a_{-}}{C}\,\sinh\!\left(C\dfrac{\pi t}{\ell}\right)\ .\end{array}\right. 
\label{eq:drive-alpha-beta-conjugacy-forms}
\end{align}

For a driving protocol composed of multiple steps as in Eq.~\eqref{eq:driven_state}, the  evolution corresponds to multiplying the associated $\mathrm{SU}(1,1)$ matrices~\eqref{eq:drive-su11-evolution-matrix}; after $N$ steps, we denote the resulting matrix entries by $(\alpha_N,\beta_N)$. This time evolution can be implemented by repeated M\"obius transformations acting on the $k$-sheeted Riemann surface associated with the $\{L_0,L_{\pm k}\}$ subalgebra~\cite{wen2018floquetconformalfieldtheory}. In particular, one finds that the point in the cylinder $w=\tau+i\,x$ is transported by the conformal map:
\begin{equation}
z \;=\; e^{\frac{2\pi}{\ell}w} , 
\qquad
z_{N}(w)\;=\; e^{\frac{2\pi }{\ell}w_{N}}
=\frac{\alpha_N\,z+\beta_N}{\beta_N^*\,z+\alpha_N^*} \ ,
\label{eq:drive-mobius-map-zn-imagepoints}
\end{equation}
which defines the ``image point'' $w_{N}$. Primary operators transform with the usual Jacobian factors, so that the driven two-point function (shown here for the chiral sector with weight $h$) in the vacuum state takes the form:
\begin{multline}
\big\langle U_N^\dagger\, \mathcal{O}(w_1,\bar w_1)\,\mathcal{O}(w_2,\bar w_2)\,U_N \big\rangle_{\rm driven}
= \\ C_{\mathcal O} \;
\prod_{j=1}^{2}\left|\alpha_N e^{\frac{2\pi i x_j}{\ell}}+\beta_N\right|^{-2h}\; 
\left[\frac{L}{\pi}\sin{\!\left(\frac{\pi}{L}\,\left|x_{N,1}-x_{N,2}\right|\right)}\right]^{-2h} 
\times \text{anti-chiral} \,,
\label{eq:drive-vacuum-two-point} 
\end{multline}
where \(w_j=i x_j\) and \(\bar w_j=-i x_j\)\footnote{Throughout this
section we evaluate correlators on an equal-time slice
$\tau_1=\tau_2=0$.}. The image points $x_{N,j}$ are obtained from Eq.~\eqref{eq:drive-mobius-map-zn-imagepoints} by evaluating $z$ at $w=ix_j$. The overall factor $C_\mathcal{O}$ collects the normalization and (if desired) the UV regulator dependence (e.g.\ powers of the short-distance cutoff $\epsilon$)

In the minimal setup, we restrict ourselves to a two-step drive per period protocol:
\begin{equation}
\ket{\Psi_N}=U^N\ket{0},\quad U=e^{-iT_1H_1}e^{-iT_0H_0} \ . \label{eq:drive-two-step-drive-protocol}
\end{equation}
Graphically
\begin{equation*}
\tikzset{every picture/.style={line width=0.75pt}} 
\begin{tikzpicture}[x=0.75pt,y=0.75pt,yscale=-1,xscale=1]
\draw    (100,127) -- (156,127) ;
\draw    (156,127) -- (156,74.61) ;
\draw    (156,74.61) -- (208,74.61) ;
\draw    (208,125.8) -- (208,74.61) ;
\draw    (208,125.8) -- (265,125.8) ;
\draw    (265,125.8) -- (264,74.61) ;
\draw    (264,74.61) -- (320,74.61) ;
\draw    (320,125.8) -- (320,74.61) ;
\draw    (320,125.8) -- (376,125.8) ;
\draw    (376,125.8) -- (375,74.61) ;
\draw    (375,74.61) -- (428,74.61) ;
\draw    (428,125.8) -- (428,74.61) ;
\draw    (428,125.8) -- (484,125.8) ;
\draw    (484,125.8) -- (483,74.61) ;
\draw    (483,74.61) -- (539,74.61) ;
\draw   (98,136.9) -- (112,139) -- (112,137.95) -- (140,137.95) -- (140,139) -- (154,136.9) -- (140,134.8) -- (140,135.85) -- (112,135.85) -- (112,134.8) -- cycle ;
\draw   (154,136.9) -- (168,139) -- (168,137.95) -- (196,137.95) -- (196,139) -- (210,136.9) -- (196,134.8) -- (196,135.85) -- (168,135.85) -- (168,134.8) -- cycle ;
\draw    (290,156.97) -- (366,156.97) ;
\draw [shift={(368,157)}, rotate = 180.88] [color={rgb, 255:red, 0; green, 0; blue, 0 }  ][line width=0.75]    (10.93,-3.29) .. controls (6.95,-1.4) and (3.31,-0.3) .. (0,0) .. controls (3.31,0.3) and (6.95,1.4) .. (10.93,3.29)   ;
\draw    (539,74.41) -- (540,125.8) ;
\draw    (540,125.8) -- (554,125.8) ;
\draw    (568,125.8) -- (582,125.8) ;
\draw    (594,125.8) -- (610,125.8) ;

\draw (109,104) node [anchor=north west][inner sep=0.75pt]   [align=left] {$\displaystyle H_0$};
\draw (171,49) node [anchor=north west][inner sep=0.75pt]   [align=left] {$H_1$};
\draw (113,147) node [anchor=north west][inner sep=0.75pt]   [align=left] {$T_0$};
\draw (169,147) node [anchor=north west][inner sep=0.75pt]   [align=left] {$T_1$};
\draw (325,158) node [anchor=north west][inner sep=0.75pt]   [align=left] {$t$};

\end{tikzpicture}
%
\end{equation*}
The driving protocol is specified by the inhomogeneous Hamiltonian
\begin{equation}
H(\theta)=\int_0 ^L\!dx\,\Big[1-\tanh(2\theta)\cos\!\Big(k\frac{2\pi x}{L}\Big)\Big]\,T_{00}(x),
\qquad k=2,3,4,\ldots,\;\theta>0,
\label{eq:drive-mobius-deformation-drive}
\end{equation}
so that $H_0$ corresponds to the uniform case $\theta=0$, \(H_0\equiv H(\theta=0)\), while turning on \(\theta\) yields the deformed step, \(H_1\equiv H(\theta\neq 0)\).

Since the system is subjected to a periodic drive protocol with period \(T=T_0+T_1\), its evolution is naturally described stroboscopically at times \(t=NT\), with $N$ the number of cycles. Notice that the Hamiltonian~\eqref{eq:drive-mobius-deformation-drive} can be expanded in terms of the generators \(\{L_0,L_{\pm k}\}\), which close into the deformed algebra \(\mathfrak{sl}^{(k)}(2,\mathbb{R})\):

\begin{equation}
H(\theta)=\frac{2\pi}{L}\Big[L_0-\tanh(2\theta)\frac{L_k+L_{-k}}{2}-\frac{c}{24} \Big]+\text{anti-chiral} \ . 
\label{eq:drive-Hdeform-virasoro-expansion}
\end{equation}
Thus, the Floquet evolution reduces to the classification of the one-cycle
evolution operator: after one period, the chiral coordinate is acted on by a single $\mathrm{SU}(1,1)$ element, and after $N$ periods, by its $N$-th power.  Denoting the one-cycle matrix by
$M\equiv M_0M_1\in \mathrm{SU}(1,1)$, the evolution after $N$ cycles is encoded in $M_N=(M_0 M_1)^N$, whose matrix
elements we denote by $(\alpha_N,\beta_N)$.

Such driven CFTs exhibit three different behaviours, commonly classified by the trace of the evolution matrix~\cite{Han:2020kwp}:
\begin{align}
\label{eq:drive-trace-classifier}
|\text{Tr}(M)| & >2 \xrightarrow \,\,\,\text{Heating Phase}\notag \\ \notag
|\text{Tr}(M)| & =2 \xrightarrow \,\,\, \text{Phase transition} \\
|\text{Tr}(M)| & <2 \xrightarrow \,\,\, \text{Non-Heating Phase}
\end{align}
Equivalently, this matches the usual classification in terms of the conjugacy class of the
corresponding $\mathfrak{sl}^{(k)}(2,\mathbb{R})$ element shown in Eq.~\eqref{eq:drive-conjugacy-classes-casimir-condition}.  For the two-step protocol used in this
work, the resulting closed-form expressions for $(\alpha_N,\beta_N)$ in the three regimes are
collected in Table~\ref{tab:drive-evolution-matrix-elements-different-phases}.

\begin{table}[t]
\centering
\resizebox{\columnwidth}{!}{%
\begin{tabular}{|c|c|}
\hline
\multicolumn{1}{|c|}{\textbf{Phase of Driven CFT}} &
\multicolumn{1}{c|}{\textbf{$\text{SU}(1,1)$ Evolution Matrix Elements}} \\ \hline
Heating Phase &
\begin{tabular}[c]{c}
$\alpha_N = (-1)^N\cosh(2N\theta)$\\
$\beta_N  = -(-1)^N\sinh(2N\theta)$
\end{tabular}
\\ \hline

Phase Transition &
\begin{tabular}[c]{c}
$\alpha_N = (-1)^N\big(1 - iN\,\tanh(2\theta)\big)$\\
$\beta_N  = -(-1)^N\big(N\,\tanh(2\theta)\big)$
\end{tabular}
\\ \hline

Non-heating Phase &
\begin{tabular}[c]{c}
$\displaystyle
(\alpha_N,\beta_N)=
\begin{cases}
(1,0), & N\equiv 0\ (\mathrm{mod}\ 4),\\
(-i\cosh 2\theta,\ i\sinh 2\theta), & N\equiv 1\ (\mathrm{mod}\ 4),\\
(-1,0), & N\equiv 2\ (\mathrm{mod}\ 4),\\
(i\cosh 2\theta,\ -i\sinh 2\theta), & N\equiv 3\ (\mathrm{mod}\ 4).
\end{cases}
$
\end{tabular}
\\ \hline

\end{tabular}%
}
\caption{Summary of the three different regimes of the two-step driving protocol~\eqref{eq:drive-two-step-drive-protocol}}
\label{tab:drive-evolution-matrix-elements-different-phases}
\end{table}
%

\subsection{Charged Moments from Composite Twist Fields in Driven CFT}

Our main goal here is to explore aspects of the symmetry-resolved entanglement entropy (SREE) introduced in Sec.~\ref{sec:sree} for a $2d$ CFT that is subjected to a drive protocol. Our discussion will not focus on specific drive protocols here, but it is general for any time evolution generated by the Hamiltonian~\eqref{eq:drive-inhomogeneous-hamiltonian}. We will then present results for the two-step protocol in Eq.~\eqref{eq:drive-two-step-drive-protocol}. 

The main ingredient to obtain the symmetry-resolved entanglement entropy~\eqref{eq:sree-symmetry-resolved-vn} is the charged moments $Z_n(\alpha)$ of the reduced density matrix $\rho_A$, introduced in Eq.~\eqref{eq:sree-Znq-and-Znalpha-definitions}. Recall that, in the replica path integral, $Z_n(\alpha)$ is represented as a two-point function of the \emph{composite twist fields} $\tau_{n,\alpha}$ and $\tilde\tau_{n,-\alpha}$, which are inserted at the
endpoints of the sub-region~$A$:
\begin{equation}
 Z_n(\alpha;t)=
\Big\langle \tau_{n,\alpha}(w_1,\bar w_1)\,\tilde\tau_{n,-\alpha}(w_2,\bar w_2)\Big\rangle_{\rm driven} \ ,
\qquad w_j=i x_j \ .
\label{eq:sree-Znalpha-driven-twist-two-point}
\end{equation}
Here, $\langle \cdot \rangle_{\rm driven}$ refers to the expectation value of the composite twist operators subjected to the Heisenberg evolution generated by a general inhomogeneous deformed CFT Hamiltonian~\eqref{eq:drive-inhomogeneous-hamiltonian} in a given state. We will choose the vacuum state because it is both simple and non-trivial\footnote{Note that, while $\{L_0, L_{\pm 1}\}$ preserves the vacuum, it evolves non-trivially under $\{L_{\pm k}\}$, for $k\ge 2$.}. Nonetheless, this framework can in principle be used to obtain the driven twist-field correlation function for an arbitrary state in the CFT\footnote{Evidently, for an arbitrary state in the CFT, the evaluation of the SREE will depend on the dynamical CFT data, since the two-point twist-field correlator now is elevated to higher-point correlation functions which involve operators that create the corresponding state from vacuum. It is certainly possible to explicitly compute everything for the free boson CFT, which is the theory mainly considered in this article. However, it is not essential for demonstrating our main point, so we will not discuss this in more detail.}.

The composite twist fields $\tau_{n, \alpha}$, $\tilde{\tau}_{n, -\alpha}$ are primaries with holomorphic weight given in Eq.~\eqref{eq:sree-twist-correlator-charged}.
For a compact boson on a circle of radius $R$, the central charge is $c=1$ and the vertex operator that implements the
${\rm U(1)}$ flux has an \emph{exactly quadratic} conformal weight~\cite{PhysRevLett.120.200602},
\begin{equation}
h_{\alpha}^V=\bar{h}_{\alpha}^V=\frac{\kappa \alpha^2}{2}
\ ,
\label{eq:sree-vertex-weight-quadratic}
\end{equation}
where $\kappa=\sqrt{2/R}$, so that
\begin{equation}
h_{n,\alpha}=
\bar h_{n,\alpha}=\frac{1}{24} \left( n - \frac{1}{n}\right)
+\frac{\kappa\alpha^2}{2n}
\ .
\label{eq:sree-hnalpha-compact-boson}
\end{equation}

We now specialize~\eqref{eq:sree-Znalpha-driven-twist-two-point} to a stroboscopic Floquet drive of $N$ cycles (we reserve $n$ for the
replica index). As seen in Eq.~\eqref{eq:drive-vacuum-two-point}, in driven CFTs, equal-time two-point correlation functions of primary fields (in vacuum or thermal state) after $N$ cycles display a
universal form. Accordingly, the two-point function~\eqref{eq:sree-Znalpha-driven-twist-two-point} of the composite twist field under  a Floquet drive takes the same form as Eq.~\eqref{eq:drive-vacuum-two-point}:
\begin{align}
Z_n(\alpha;N)
=
C_{n,\alpha}\;\prod_{j=1}^{2}
\left|\alpha_N e^{\frac{2\pi i x_j}{\ell}}+\beta_N\right|^{-2h_{n,\alpha}}\;
\left[\frac{L}{\pi\epsilon}\sin{\!\Big(\frac{\pi}{L}\,|x_{N,1}-x_{N,2}|\Big)}\right]^{-2h_{n,\alpha}} \notag \\  \times
\left[\text{anti-chiral part with }h_{n,\alpha}\to \bar h_{n,\alpha}\right].
\label{eq:sree-Znalpha-driven-universal-form}
\end{align}
Here, $(\alpha_N,\beta_N)$ are the $\mathrm{SU}(1,1)$ matrix elements~\eqref{eq:drive-su11-evolution-matrix} after $N$ cycles and $x_{N,j}$ are the transported points defined implicitly via the
M\"obius action~\eqref{eq:drive-mobius-map-zn-imagepoints}. The correlator must be regulated at the endpoints of $A$ by a UV cutoff $\epsilon$.
If we package all the $N$-dependence into the dimensionless driven scale,
\begin{equation}
\mathcal{D}_N \;\equiv\;
\frac{L}{\pi\,\epsilon}\,
\sin{\!\Big(\frac{\pi}{L}\,|x_{N,1}-x_{N,2}|\Big)}\,
\left|\alpha_N e^{\frac{2\pi i x_1}{\ell}}+\beta_N\right|\,
\left|\alpha_N e^{\frac{2\pi i x_2}{\ell}}+\beta_N\right| \ ,
\label{eq:sree-driven-scale-DN}
\end{equation}
and taking into account that $\bar{h}_{n, \alpha}=h_{n,\alpha}$, then Eq.~\eqref{eq:sree-Znalpha-driven-universal-form} may be written compactly as:
\begin{equation}
Z_n(\alpha;N)=C_{n,\alpha}\;\mathcal{D}_N^{-4h_{n,\alpha}} \ .
\label{eq:sree-Znalpha-compact-DN-form}
\end{equation}
The split between the explicit cutoff $\epsilon$ inside $\mathcal{D}_N$ and the constant $C_{n,\alpha}$
is scheme dependent; only their combination in Eq.~\eqref{eq:sree-Znalpha-compact-DN-form} is meaningful. If we now insert the \emph{exact} dimension of the composite twist field for a compact boson~\eqref{eq:sree-hnalpha-compact-boson}, we have:
\begin{align}
Z_n(\alpha;N)
&=C_{n,\alpha}\,\mathcal{D}_N^{-4\left(h_{n}+\frac{\kappa\alpha^2}{2n}\right)}.
\nonumber
\label{eq:sree-Znalpha-factorized-dim-inserted}
\end{align}

This expression can be rewritten as:
\begin{equation}
 Z_n(\alpha;N)
=
C_{n,\alpha}\;\mathcal{D}_N^{-4h_{n}}\;
\exp\!\left[
-\frac{2\kappa\alpha^2}{n}\log \mathcal{D}_N
\right] \ .
\label{eq:sree-Znalpha-gaussian-logDN}
\end{equation}
It is useful to compare it with the uncharged moment:
\begin{equation}
 Z_n(0;N)=C_{n,0}\,\mathcal{D}_N^{-4h_{n}} \ ,
\label{eq:sree-Zn0-uncharged-moment}
\end{equation}
which follows from \eqref{eq:sree-Znalpha-gaussian-logDN} by setting $\alpha=0$.
Taking their ratio, we find:
\begin{align}
\frac{ Z_n(\alpha;N)}{ Z_n(0;N)}
&=
\frac{C_{n,\alpha}}{C_{n,0}}\,
\exp\!\left[
-\frac{2\kappa\alpha^2}{n}\log \mathcal{D}_N
\right] \ .
\label{eq:sree-ratio-Znalpha-Zn0}
\end{align}
The corresponding R\'enyi-$n$ entropy and entanglement entropy are related to the uncharged moment~\eqref{eq:sree-Zn0-uncharged-moment} by~\cite{cc-04, wen2018floquetconformalfieldtheory}:
\begin{align}
    S^{(n)}_A(N)&=\frac{1}{1-n}\log{(Z_n(0;N))}=\frac{n+1}{6n}\log(\mathcal{D}_N),\\
    S_A(N)&=-Z_n'(0;N)\Bigg|_{n=1}=\frac{1}{3}\log{(\mathcal{D}_N)}\,.\label{eq:sree-renyi-and-ee-from-Zn0}
\end{align}
This is the result obtained in Ref.~\cite{wen2018floquetconformalfieldtheory} for the evolution of the entanglement entropy under a stroboscopic Floquet drive after $N$ cycles starting from the ground state of the CFT. We will derive an analogous expression for the symmetry-resolved entanglement entropies.

Coming back to the charged moments~\eqref{eq:sree-ratio-Znalpha-Zn0}, the remaining flux dependence which is not fixed by conformal symmetry is encoded in the
(time-independent) normalization $C_{n,\alpha}$. This normalization constant does not depend on the drive protocol, however, it depends on the R\'enyi index $n$ as well as the flux $\alpha$. We will assume that the normalization has a smooth limit in the vanishing flux parameter regime. This allows us to perform a Taylor expansion of $C_{n,\alpha}$ in $\alpha$. As done in the vacuum state with no driving~\cite{bonsignori19}, we parametrize the small-flux behavior by an exactly quadratic dependence:\footnote{We are implicitly assuming a symmetry under $\alpha \to - \alpha$, since the $\alpha$-dependence of the dimension of the composite twist fields obey this symmetry.}
\begin{equation}
\frac{C_{n,\alpha}}{C_{n,0}}
=
\exp\!\left(\gamma(n)\alpha^2+\mathcal{O}(\alpha^4)\right) \ ,
\qquad
\gamma(n)\ \text{independent of }N \ .
\label{eq:sree-normalization-small-flux}
\end{equation}
 Inserting the expansion in Eq.~\eqref{eq:sree-normalization-small-flux} into \eqref{eq:sree-ratio-Znalpha-Zn0} yields:

\begin{equation}
 Z_n(\alpha;N)= Z_n(0;N)\,\exp\!\big[-B_n(N)\,\alpha^2\big] \ ,
\label{eq:sree-Znalpha-gaussian-Bn}
\end{equation}
with:
\begin{equation}
B_n(N)
=\frac{2\kappa}{n}\log \mathcal{D}_N-\gamma(n)\,.
\label{eq:sree-Znalpha-Bn}
\end{equation}
Clearly, the above result holds under the Gaussian approximation~\eqref{eq:sree-normalization-small-flux}, in which the normalization constant depends only quadratically on $\alpha$. In principle, it is possible to assume a more general expansion, including non-Gaussianity in the flux, and subsequently obtain a perturbative correction for the charged moments around the Gaussian result. 

The result in Eq.~\eqref{eq:sree-Znalpha-gaussian-Bn} is formally similar to that obtained in Refs.~\cite{PhysRevLett.120.200602, bonsignori19} for the charged moments in the ground state of the compact boson, upon replacing the subsystem chord length
$L/\pi
\sin\left(\pi|x_1-x_2|/L\right)$ by the term $\mathcal{D}_N$ introduced in Eq.~\eqref{eq:sree-driven-scale-DN}, which encodes the dependence on the driving protocol and plays the role of an effective subsystem length.
Eq.~\eqref{eq:sree-Znalpha-gaussian-Bn} further extends the results of Ref.~\cite{Chen:2025fzm} for the evolution of charged moments under quenches with the inhomogeneous Hamiltonians~\eqref{eq:drive-Hdeform-virasoro-expansion} to the case of a generic driving protocol of the form~\eqref{eq:driven_state}.

\subsection{SREE in Different Phases of Driven CFT}

To derive the symmetry-resolved entanglement entropies~\eqref{eq:sree-renyi-via-Znq}, we need to determine the fixed-charge moments $\mathcal Z_n(q;N)$. These are obtained by Fourier transforming the charged moments $Z_n(\alpha;N)$ with respect to the flux parameter $\alpha$:
\begin{equation}
\label{eq:sree-fixedcharge-fourier}
\mathcal Z_n(q;N)
= \int_{-\pi}^{\pi}\! \frac{d\alpha}{2\pi}\;
Z_n(\alpha;N)\,e^{-iq\alpha}\,.
\end{equation}
Within the Gaussian approximation~\eqref{eq:sree-normalization-small-flux} for the normalization term $C_{n, \alpha}$,
$Z_n(\alpha;N)$ is given by Eq.~\eqref{eq:sree-Znalpha-gaussian-Bn}.
Inserting it in Eq.~\eqref{eq:sree-fixedcharge-fourier}, the $\alpha$-integral yields:
\begin{equation}
\label{eq:sree-fixedcharge-erf-exact}
\mathcal Z_n(q;N)
=\frac{Z_n(0;N)}{4\,\sqrt{\pi B_n(N)}}\,
\exp\!\left[-\frac{q^2}{4B_n(N)}\right]\,
E\!\left[B_n(N);q\right] \,,
\end{equation}
with the error-function combination:
\begin{equation}
\label{eq:sree-E-definition}
E\left[B_n(N);q\right]
\equiv
\operatorname{erf}\left(\pi\sqrt{B_n(N)}+\frac{i\,q}{2\sqrt{B_n(N)}}\right)
+\operatorname{erf}\left(\pi\sqrt{B_n(N)}-\frac{i\,q}{2\sqrt{B_n(N)}}\right)\,.
\end{equation}

For \(n=1\), the uncharged moment reduces to the trace of the reduced density matrix,
\(Z_1(0;N)=\Tr(\rho_A)=1\).

We will also require the derivative of $B_1(N)$ with respect to the replica index at \(n=1\), which we denote by
\begin{equation}
\label{eq:sree-Bcal-definition}
\mathcal{B}(N)\;\coloneqq\;\left.\frac{\partial B_n(N)}{\partial n}\right|_{n=1}\,.
\end{equation}

In the regime where the effective driven scale is large, {\it i.e.}~$\log\mathcal{D}_N\gg 1$, the width
$B_1(N)\propto \log\mathcal{D}_N$ is large. Then $\pi\sqrt{B_n(N)}\gg 1$ and the error functions in
Eq.~\eqref{eq:sree-fixedcharge-erf-exact} approach unity (up to exponentially small corrections in $B_n$). Explicitly, the error functions can be expanded in the form\footnote{For large $|z|$ (with $|\arg z|<\tfrac{3\pi}{4}$), we have the standard asymptotic form of the complementary error function
$\operatorname{erfc}(z)\equiv 1-\operatorname{erf}(z)$:
\begin{eqnarray}
\operatorname{erfc}(z)\sim \frac{e^{-z^{2}}}{\sqrt{\pi}\,z}\Big(1-\frac{1}{2z^{2}}+\frac{3}{(2z^{2})^{2}}-\frac{15}{(2z^{2})^{3}}+\cdots\Big)
=\frac{e^{-z^{2}}}{\sqrt{\pi}\,z}\sum_{n=0}^{\infty}\frac{(-1)^n(2n-1)!!}{(2z^{2})^{n}} \ , \label{erf_expand}
\end{eqnarray}
so that $\operatorname{erf}(z)=1-\operatorname{erfc}(z)$.}
:
\begin{equation}
\erf\!\left(\pi\sqrt{B_n}+\frac{i q}{2\sqrt{B_n}}\right)+
\erf\!\left(\pi\sqrt{B_n}-\frac{i q}{2\sqrt{B_n}}\right)
=2+\mathcal{O}\!\left(e^{-\pi^2 B_n}\right) \ .
\label{eq:sree-erf-large-Bn-asymptotic}
\end{equation}
Keeping only the leading term, the fixed-charge moments reduce to:
\begin{align}
\mathcal Z_n(q;N)\simeq\;\frac{Z_n(0;N)}{2\,\sqrt{\pi B_n(N)}}\,
\exp\left[-\frac{q^2}{4B_n(N)}\right],\notag \\
\mathcal Z_1(q;N)\simeq\frac{1}{2\sqrt{\pi B_1(N)}}\,
\exp\left[-\frac{q^2}{4B_1(N)}\right] \ .
\label{eq:sree-fixed-charge-moment-gaussian-approx}
\end{align}
Now, from Eq.~\eqref{eq:sree-renyi-via-Znq}, we obtain the symmetry resolved R\'enyi entropy:
\begin{align}
     S_A^{(n)}(q,N) &=S_A^{(n)}(N)-\log{2}+\frac{1}{2(1-n)}\log{\left(\frac{B_1^{n}}{B_n}\right)}-\frac{q^2}{4(1-n)}\left(\frac{n}{B_1}-\frac{1}{B_n}\right) \label{eq:sree-renyi-fixed-charge}\ ,
\end{align}
where $S_A^{(n)}(N)$ is the total R\'enyi entropy in Eq.~\eqref{eq:sree-renyi-and-ee-from-Zn0}. Taking the $n\rightarrow1$ limit in this expression, we obtain the symmetry resolved entanglement entropy:
\begin{equation}
    S_A(q,N) =S_A(N)-\frac{1}{2}\log{(4\pi \,B_1(N))}+\frac{\mathcal{B}(N)}{2B_1(N)}-\frac{\mathcal{B}(N)+B_1(N)}{4B_1(N)^2}q^2 \label{eq:sree-vn-fixed-charge}\ .
\end{equation}
The first three terms in this expression are independent of the charge $q$ and contain the entire leading UV-sensitive contribution; the {\it only} charge dependence appears through the quadratic correction $\propto q^2/B_1(N)^2$, which is suppressed at large $B_1(N)$. 
In that regime, $S_A(q,N)$ becomes effectively $q$-independent to leading order, i.e. the entanglement is (approximately) {\it equipartitioned} among all the charge sectors~\cite{xavier2018, bonsignori19,Chen:2025fzm}. 
In the decomposition in Eq.~\eqref{eq:sree-entropy-decomposition-config-number}, this means that essentially all nontrivial $q$-dependence resides in the classical weights $p_q$ (charge fluctuations), while the entanglement {\it within} each fixed-charge sector is the same up to corrections suppressed by inverse powers of $B_1(N)$. 

To further analyze the result in Eq.~\eqref{eq:sree-vn-fixed-charge}, we consider single-interval subsystems whose end-points are chosen to coincide with the spatial deformation wavelength
$\ell=L/k$, such as:
\begin{equation}
A=\Big[(p-\tfrac12)\ell,\,(p+j-\tfrac12)\ell\Big]
\,\,\, \text{or} \,\,\,
A=\big[p\ell,\,(p+j)\ell\big],
\,\,\, p,j\in\mathbb{Z},\ \ 1\le j<k.
\end{equation}
Under the $k$-fold uniformization \(z=e^{\frac{2\pi k}{L}w}\) (at equal Euclidean time \(w=ix\)),
the two endpoints map to the same point in the \(z\)-plane but lie on different sheets of the
$k$-sheeted cover, separated by \(j\) layers; upon passing to the single-valued coordinate
\(\zeta=z^{1/k}\), their separation becomes a purely
geometric factor \(\big(2\sin\frac{\pi j}{k}\big)^{-4h}\) in the twist-field correlation~\eqref{eq:drive-vacuum-two-point}. Physically, restricting to such ``cell-aligned'' intervals is natural because, under an $\mathfrak{sl}^{(k)}(2,\mathbb{R})$ driving, the expectation value of the energy-momentum density develops sharp, localized peaks in real space~\cite{Das:2024lra,Wen:2022pyj}, whose locations are determined by the stroboscopic M\"obius data. For an interval $A=[x_1,x_2]$, the dominant stroboscopic time-dependence of the entanglement comes from the conformal Jacobians evaluated at the end-points $x_1,x_2$, so the qualitative behavior is controlled by the relative position of these end-points with respect to the stress-tensor peaks. If a peak coincides with an end-point, the degrees of freedom responsible for inter-region entanglement effectively pile up at the boundary and the entropy can \emph{decrease} with time (in special protocols even linearly), despite the fact that the total energy still grows. Away from this fine-tuned situation --- i.e., for generic end-point positions that do not sit on the peak centers --- the stress-tensor peaks lie in the \emph{interior} of the chosen interval, while the end-points fall between the peak locations; in this case, the conformal map effectively stretches the separation between the end-points and the entanglement increases. In the heating phase, this typically yields linear growth with the number of drive cycles, whereas at phase transition the growth is only logarithmic~\cite{Wen:2020wee,Wen:2021mlv,Fan_2021}.

\begin{figure}[!htbp]
    \centering
    \vspace{2mm}
    {\includegraphics[width=0.49\textwidth]{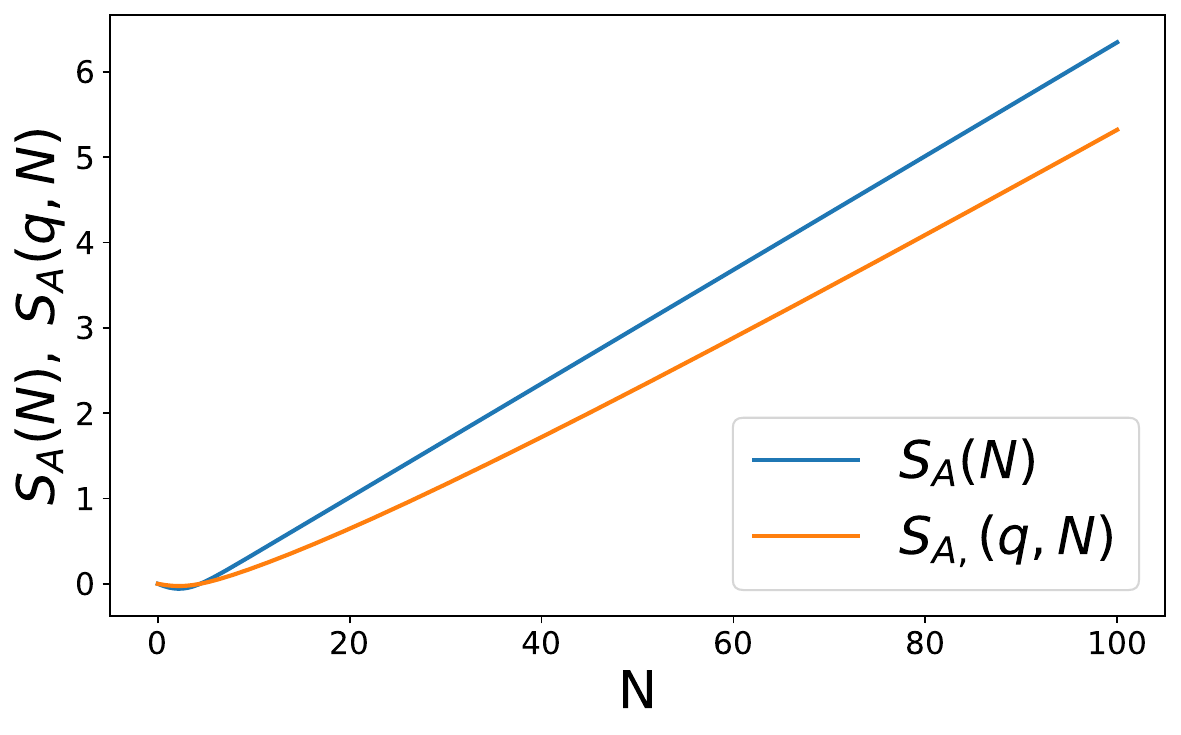}}
    \hfill
    {\includegraphics[width=0.49\textwidth]{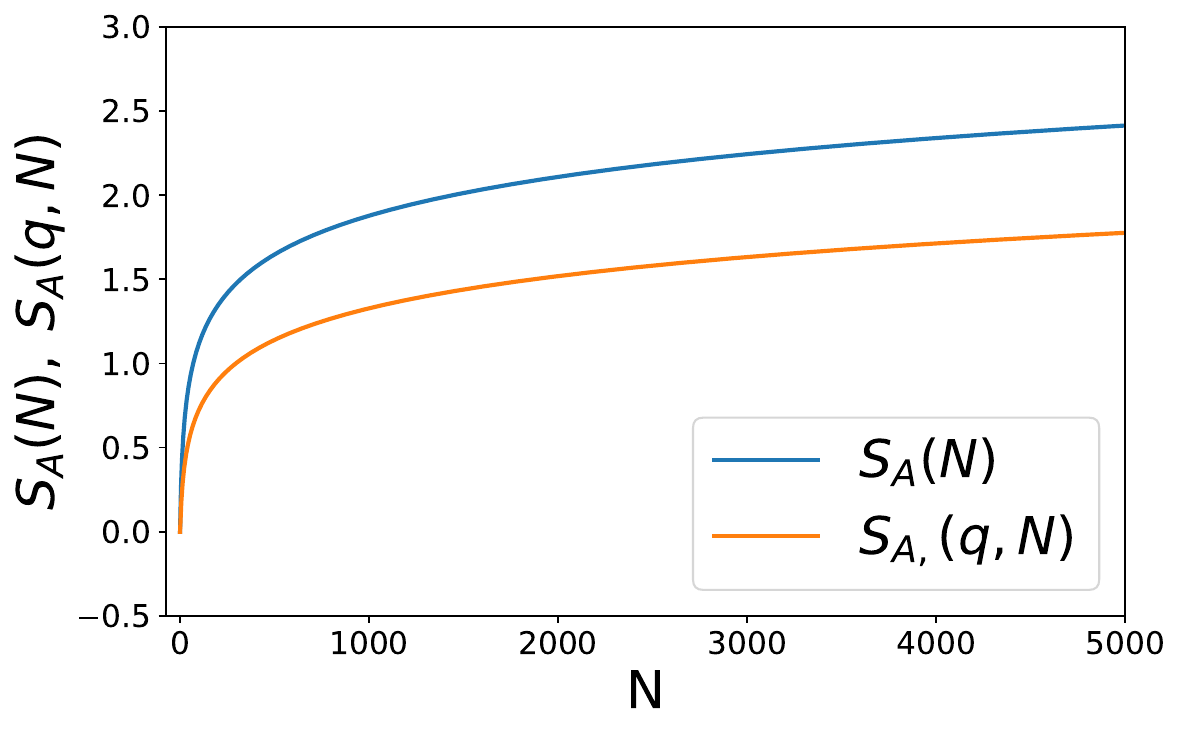}}

    \vspace{2mm}
    {\includegraphics[width=0.6\textwidth]{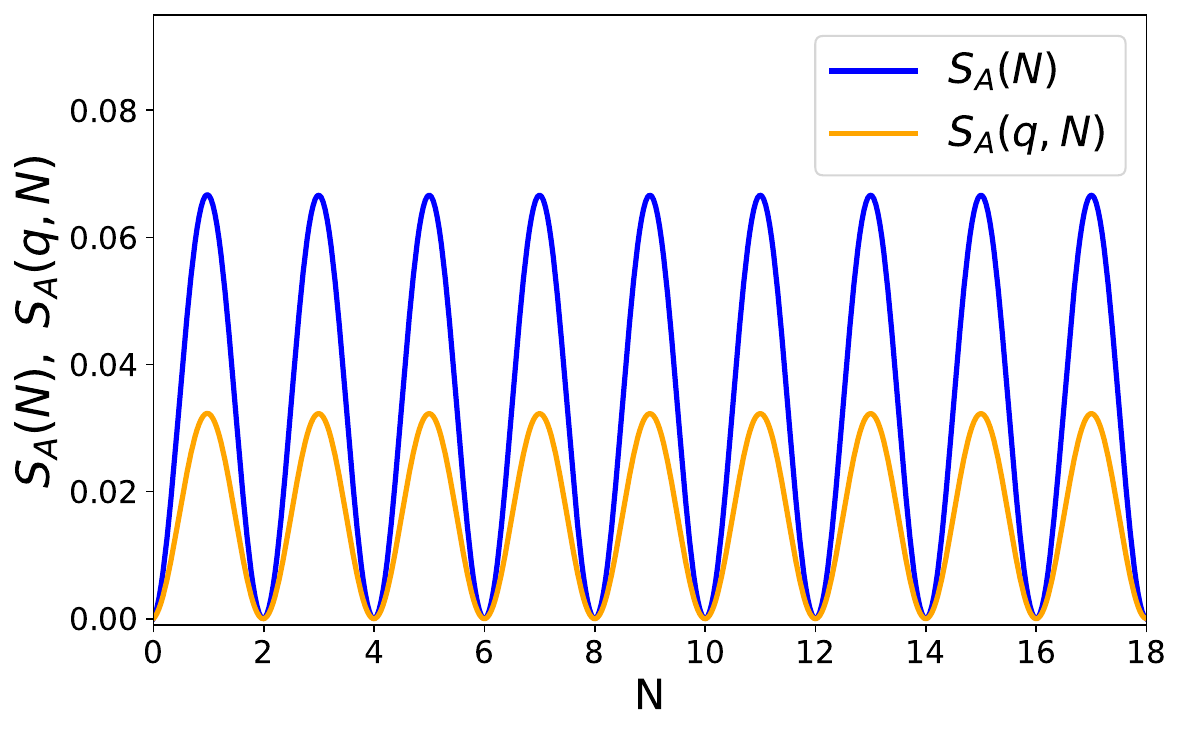}}
    \caption{Evolution of the symmetry-resolved entanglement entropy, $S_A(q;N)$, as a function of the number of drive cycles $N$ in the heating phase (top-left), at the phase transition (top-right), and in the non-heating phase (bottom) of the driving protocol~\eqref{eq:drive-two-step-drive-protocol}. We take $k=3$, $q=2$, $L=1000$, and the driving parameter $\theta=0.1$. As a benchmark, we also show the evolution of the total von Neumann entanglement entropy, $S_A(N)$, as a function of the number of drive cycles $N$.}
    \label{fig:sree-driven-cft}
\end{figure}
%

We therefore restrict to ``cell-aligned'' intervals whose end-points are placed on this grid, so that the subsystem
length is quantized in units of $\ell$,
\begin{equation}
l \equiv |x_2-x_1| = j\,\ell = j\,\frac{L}{k},\qquad k,j\in\mathbb{Z},\ \ 1\le j<k .
\end{equation}
With this choice, the conformal distances entering the twist-field correlator (and hence the EE/SREE) are controlled by the driven scale $l=jL/k$ rather than directly by $L$. Importantly, even in the thermodynamic limit $L\to\infty$, the interval length $l$ need not become asymptotically large if $k$ is taken large simultaneously. For instance, one may keep $L/k$ fixed, or more generally allow $k$ to scale with $L$, so that for fixed UV cutoff $\epsilon$ and fixed $j$ the quantity $\log(l/\epsilon)=\log(jL/(k\epsilon))$ need not be parametrically large. In this regime, the charge-dependent term $\propto q^2/B_1$, with $B_1\sim \log \mathcal{D}_N$, in Eq.~\eqref{eq:sree-vn-fixed-charge} is no longer strongly suppressed, and visible deviations from charge equipartition can arise. 

In Fig.~\ref{fig:sree-driven-cft}, we plot the time evolution under the two-step Floquet protocol of Eq.~\eqref{eq:drive-two-step-drive-protocol} for both the total entanglement entropy and the symmetry-resolved entanglement entropy in the heating phase (top-left panel), at the phase transition (top-right panel) and in the non-heating phase (bottom panel). We observe that, for a large number of driving cycles, the symmetry-resolved entropy behaves similarly to the total entropy: it grows linearly in the heating phase, logarithmically at the phase transition, and oscillates in the non-heating phase.

\subsection{Breakdown of Charge Equipartition}

Let us now return to the exact fixed-charge moment \eqref{eq:sree-fixedcharge-erf-exact} and keep the expression exact, without invoking any approximation at this stage. The symmetry resolved entanglement entropy then follows from its definition:
\begin{equation}
    S_A(q;N) =S_A(N)-\frac{1}{2}\log{\left(\frac{16\pi \,B_1}{E^2[B_1,q]}\right)}+\frac{\mathcal{B}}{2B_1}-\frac{\mathcal{B}+B_1}{4B_1^2}q^2-\mathcal{B}\frac{\mathcal{E}[B_1,q]}{E[B_1,q]} \,, \label{eq:sree-vn-fixedcharge-erf-exact}
\end{equation}
where \(\mathcal{B}(N)\) is defined in \eqref{eq:sree-Bcal-definition}, and
\begin{equation}
\label{eq:sree-Ecal-definition}
\mathcal{E}\!\left[B_1(N);q\right]\;=\;
\left.\frac{\partial}{\partial n}\,E\!\left[B_n(N);q\right]\right|_{n=1}\,.
\end{equation}
Equation~\eqref{eq:sree-vn-fixedcharge-erf-exact} makes the breaking of charge equipartition manifest whenever $B_1(N)$ is not parametrically large: the $q$-dependence is controlled explicitly by the terms proportional to $q^2$ and by the ratio $\mathcal{E}/E$. In the opposite regime $B_1(N)\gg 1$, the error-function factor approaches
$E[B_1(N),q]\simeq 2$ , so that Eq.~\eqref{eq:sree-vn-fixedcharge-erf-exact} reduces to Eq.~\eqref{eq:sree-vn-fixed-charge}.

To analyze with more detail the crossover between these regimes, we can use in Eq.~\eqref{eq:sree-fixedcharge-erf-exact} the expansion of the error function in Eq.~(\ref{erf_expand}), which is valid for any value of $B_1(N)$. This yields:
\begin{equation}
\label{eq:sree-Z1q-asymptotic-generalB}
\mathcal Z_1(q)
=
\sqrt{\frac{\pi}{B_1(N)}}\;e^{-\,\frac{q^2}{4B_1(N)}}
\;-\;
\frac{4\pi B_1(N)\,e^{-B_1(N)\pi^2}}{q^2}\cos(q\pi)
+\mathcal{O}\!\left(\frac{1}{|q|^4}\right) \ .
\end{equation}
For integer $q$ one has $\cos(q\pi)=(-1)^q$, making the second term above explicitly oscillatory. This oscillation originates from the phase $e^{-iq\alpha}$ in Eq.~\eqref{eq:sree-fixedcharge-fourier}
evaluated at the flux integration endpoints $\alpha=\pm\pi$, i.e.\ $e^{\pm iq\pi}$.
The first term in \eqref{eq:sree-Z1q-asymptotic-generalB} is exponentially suppressed as
$e^{-q^2/(4B_1)}$, while the second term comes with the factor $e^{-B_1\pi^2}$.
A parametric crossover occurs when these exponentials are comparable:
\begin{equation}
\label{eq:sree-charge-crossover-scale}
e^{-\,\frac{q^2}{4B_1(N)}}\sim e^{-B_1(N)\pi^2}
\quad\Longleftrightarrow\quad
|q|\sim 2\pi B_1(N) \ .
\end{equation}
Thus, for $|q|\ll 2\pi B_1(N)$, the bulk Gaussian term dominates and $\mathcal{Z}_1(q)$ is well
approximated by \eqref{eq:sree-fixed-charge-moment-gaussian-approx}. In contrast, for $|q|\gtrsim 2\pi B_1(N)$ the bulk term is suppressed compared to the flux endpoint contribution. The leading behaviour is then governed by
the second term in Eq.~\eqref{eq:sree-Z1q-asymptotic-generalB} which exhibits the characteristic $(-1)^q$ oscillation. Therefore, even when $B_1(N)$ is \emph{not} parametrically large, as long as we restrict to the
charge window:
\begin{equation}
|q|\ll 2\pi B_1(N) \ ,
\label{eq:sree-gaussian-charge-window}
\end{equation}
the bulk term dominates and $\mathcal{Z}_1(q;N)$ is well-approximated by the Gaussian form:
\begin{equation}
\mathcal Z_1(q;N)\simeq \frac{\sqrt{\pi}}{2\pi\sqrt{B_1(N)}}\exp\!\Big(-\frac{q^2}{4B_1(N)}\Big) \ .
\label{eq:sree-Z1q-gaussian-window}
\end{equation}
In this regime the symmetry-resolved entanglement entropy reduces to the same functional structure as the Gaussian result:
\begin{equation}
 S_A(q;N) =S_A(N)-\frac{1}{2}\log{(4\pi \,B_1)}+\frac{\mathcal{B}}{2B_1}-\frac{\mathcal{B}+B_1}{4B_1^2}q^2 \ .
 \label{eq:sree-vn-fixedcharge-gaussian-window}
\end{equation}
As far as equipartition is concerned, the interpretation changes when $B_1(N)$ is only moderate. In the large-$B_1$ limit, the $q^2$ term is suppressed by $1/B_1$ and thus becomes negligible for fixed $q$, leading to approximate charge equipartition across sectors. In contrast, when $B_1(N)$ is not large, the ratio $q^2/B_1(N)$ need not be small and the $q$-dependence in Eq.~\eqref{eq:sree-vn-fixedcharge-gaussian-window} can be sizable. In this sense,
equipartition is explicitly broken, even though $\mathcal Z_1(q;N)$ remains well-controlled by the Gaussian approximation~\eqref{eq:sree-gaussian-charge-window}.

This comes with a clear trade-off: the range of charge sectors for which the Gaussian answer is reliable shrinks as $B_1(N)$ decreases, because the condition \eqref{eq:sree-gaussian-charge-window} becomes more restrictive. Equivalently, for smaller $B_1(N)$ one reaches the crossover \eqref{eq:sree-charge-crossover-scale} already at relatively small integers $q$, beyond which cutoff effects (the oscillatory term in \eqref{eq:sree-Z1q-asymptotic-generalB}) and, more
generally, the exact error-function dependence become important. A fully controlled analysis outside the window
\eqref{eq:sree-gaussian-charge-window} should then revert to the exact SREE expression in terms of the error-function factor \eqref{eq:sree-vn-fixedcharge-erf-exact},
where the breaking of equipartition is manifest through both the explicit $q^2$ term and the $q$-dependence
encoded in the error-function ratios.
\begin{figure}[t]  
    \centering
     \vspace{2mm}
    \includegraphics[width=0.49\linewidth]{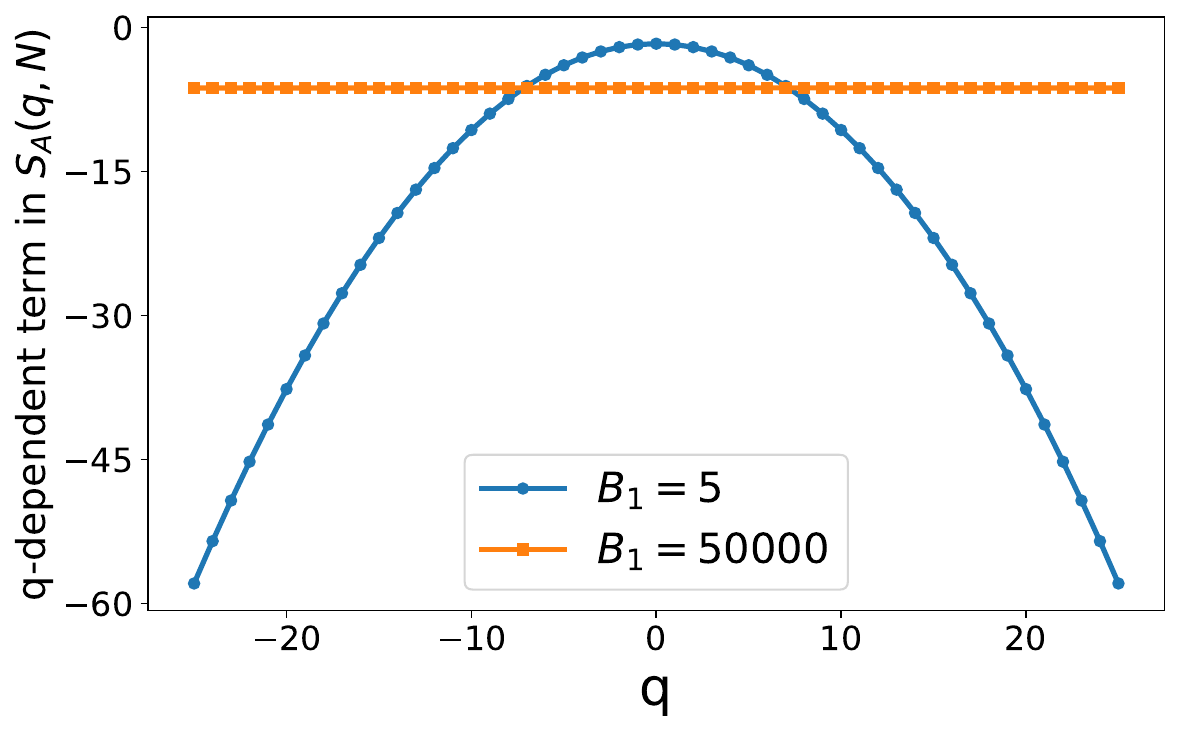}
    \hfill
    {\includegraphics[width=0.49\textwidth]{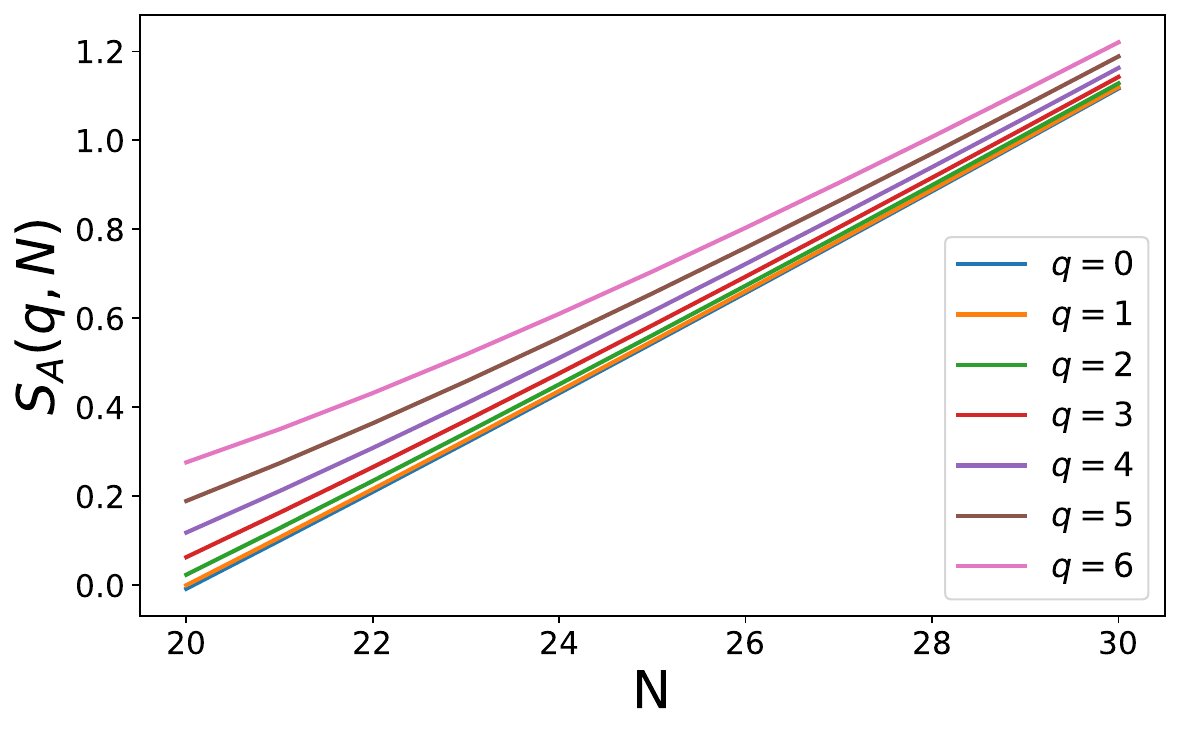}}
\vspace{2mm}
    {\includegraphics[width=0.49\textwidth]{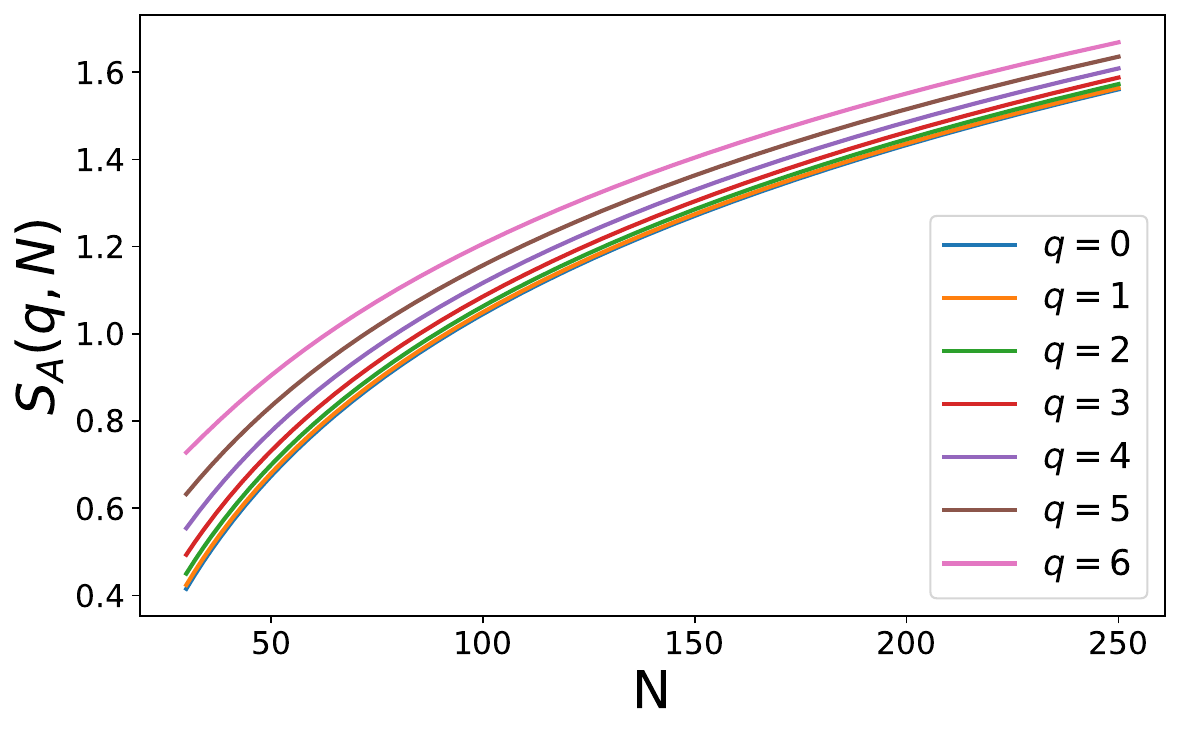}}
    \hfill
    {\includegraphics[width=0.49\textwidth]{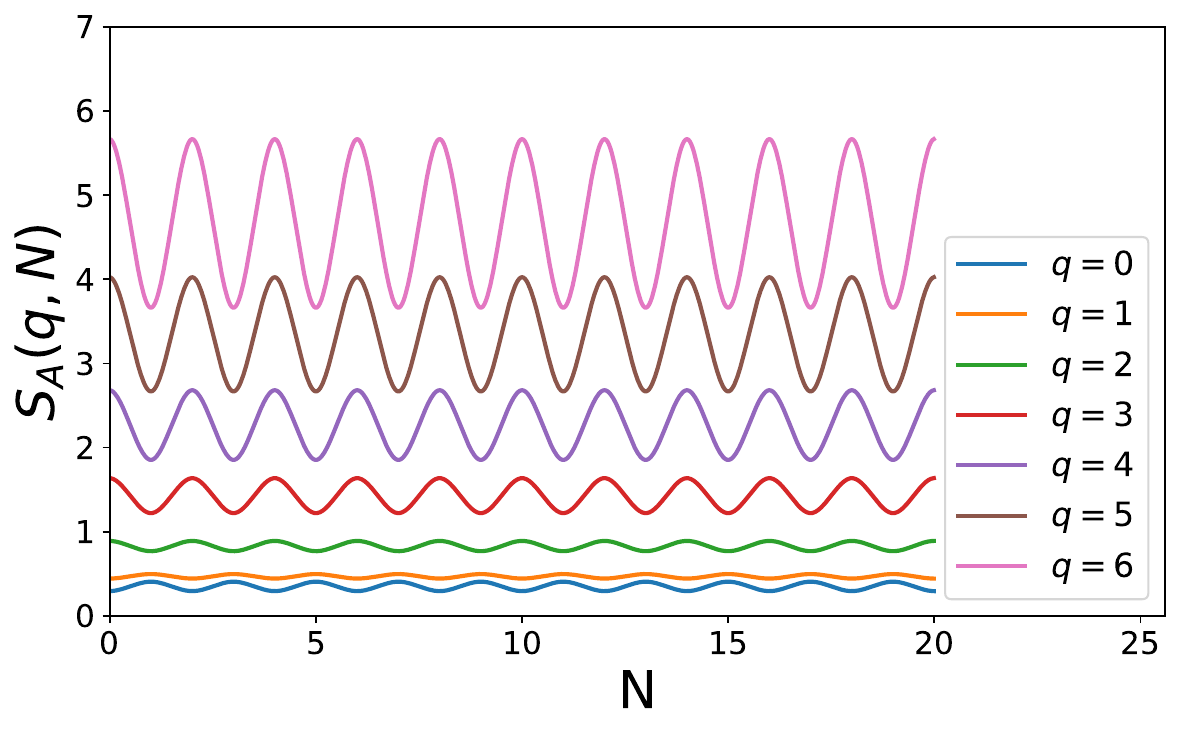}}
    \caption{
The top-left panel shows the $q$-dependent contribution to the SREE, shown as a function of charge $q$ for a representative moderate width ($B_1=5$, blue dots) and a large width ($B_1=5\times10^{4}$, orange squares). The top-right, bottom-left, and bottom-right panels display stroboscopic evolution of SREE for different charge sectors \(q\) in the heating, phase-transition, and non-heating phases of the two-step driving protocol~\eqref{eq:drive-two-step-drive-protocol}, respectively.  For these three figures, we set \(L=1200\), \(k=120\), and the driving parameter \(\theta=0.1\), such that the effective scale \(\ell=L/k\) is sufficiently small to make the breakdown of charge equipartition clearly visible.
}
    \label{fig:sree-equipartition-breaking}
\end{figure}

A further implication of the crossover \eqref{eq:sree-charge-crossover-scale} is that, once one probes charges $|q|\gtrsim 2\pi B_1(N)$,
the finite integration range $\alpha\in[-\pi,\pi]$ can no longer be ignored and the large-$|q|$
endpoint contribution dominates the Fourier coefficient, which alternates in sign as a function of even and odd charges. In particular, in the cutoff-dominated regime where the endpoint term
overwhelms the bulk Gaussian, one finds schematically:
\begin{equation}
\mathcal Z_1(q;N)\;\sim\;-\frac{\mathcal{A}(B_1)}{q^2}\,(-1)^q \ ,
\qquad |q|\gtrsim 2\pi B_1(N) \ ,\label{eq:sree-Z1q-endpoint-oscillatory}
\end{equation}
so that $\mathcal Z_1(q)$ changes sign for every step in $q$. Note that, since a physical partition function is positive, this implies the existence of a maximum charge $q_{\rm max} \sim 2\pi B_1(N)$ beyond which the results are not valid. In lattice models, such as spin chains, the finite number of degrees of freedom naturally introduces a maximum charge, but not always, {\it e.g.}, bosonic systems, where occupation numbers are unbounded.

To further illustrate the breakdown of charge equipartition, in Fig.~\ref{fig:sree-equipartition-breaking} we first show, in the top-left panel, the \(q\)-dependent contribution to the SREE for representative moderate and large values of \(B_1\). For moderate \(B_1\), the dependence on the charge sector is clearly visible and is approximately quadratic in \(q\), indicating a pronounced departure from equipartition. By contrast, for large \(B_1\) this \(q\)-dependence is strongly suppressed and the curve becomes almost level, consistent with the restoration of charge equipartition in the large-\(B_1\) regime. 
In the remaining three panels, we plot the symmetry-resolved entanglement entropy \(S_A(q;N)\) for different charge sectors \(q\) in the heating, phase-transition, and non-heating regimes of the two-step driving protocol~\eqref{eq:drive-two-step-drive-protocol}, shown in the top-right, bottom-left, and bottom-right panels, respectively. For these plots, we choose the effective scale \(L/k\) to be sufficiently small. In this regime, the width \(B_1(N)\) is not parametrically large, and the charge-dependent corrections therefore are still significant. As a result, the curves corresponding to different charge sectors remain visibly separated, showing explicitly that the SREE retains a nontrivial \(q\)-dependence and does not exhibit equipartition. The detailed dependence on the number of drive cycles \(N\) remains phase specific, as discussed earlier. After sufficiently many drive cycles, \(B_1(N)\) can become parametrically large, leading to an asymptotic restoration of charge equipartition. For sufficiently small \(L/k\), such asymptotic restoration occurs only in the heating phase and at the phase transition, where \(B_1(N)\) grows without bound with the number of drive cycles. In the non-heating phase, \(B_1(N)\) remains bounded and oscillatory, so equipartition is not restored dynamically. Moreover, the restoration is parametrically slower at the phase transition than in the heating phase, since \(B_1(N)\) grows only logarithmically with \(N\) at phase transition, whereas it grows linearly in the heating regime. This behavior is also evident from Fig.~\ref{fig:sree-equipartition-breaking}.

\subsection{Number (Shannon) Entropy in Different Phases of Driven CFT}\label{sec:number-entropy}

As reviewed in Sec.~\ref{sec:sree} (see in particular Eq.~\eqref{eq:sree-entropy-decomposition-config-number}),
the number entropy $S_{\rm num}$ captures the entanglement associated with fluctuations of the conserved
$U(1)$ charge within the subsystem. Equivalently, it is the Shannon entropy of the fixed-charge probability
distribution in $A$ at stroboscopic time $N$:
\begin{equation}
S_{\rm num}(N)\;\equiv\;-\sum_{q} p_q(N)\,\log p_q(N)\,.
\label{eq:sree-number-entropy-definition}
\end{equation}
Physically, $p_q(N)$ is the probability that a projective measurement of $Q_A$ at stroboscopic time
$N$ yields charge $q$. Hence $S_{\rm num}$ quantifies the \emph{classical uncertainty} in the charge
sector of $A$, i.e.\ the amount of entanglement that is generated purely by fluctuations of the
conserved quantity between $A$ and $B$. In particular, $S_{\rm num}=0$ when the subsystem charge is
sharp, \textit{i.e.} only one $p_q$ is nonzero, whereas broader charge distributions lead to larger number entropy.

In our notation, the sector weight is given by the $n=1$ fixed-charge moment,
\begin{equation}
p_q(N)=\mathcal Z_1(q;N)\,.
\label{eq:sree-sector-probability}
\end{equation}
In the regime where the approximation~\eqref{eq:sree-fixed-charge-moment-gaussian-approx} is reliable (equivalently, when the width parameter
$B_1(N)$ is sufficiently large), the discrete sum over integer charges can be approximated by an
integral and the charge distribution is well described by a normalized Gaussian. This form makes the physical meaning of $B_1(N)$ transparent: it sets the variance of the charge
distribution,
\begin{equation}
\mathrm{Var}(Q_A)\;\equiv\;\sum_q q^2\,p_q(N)\;\simeq\;\int_{-\infty}^{\infty}\!dq\;q^2\,\mathcal Z_1(q;N)
\;=\;2\,B_1(N),
\label{eq:sree-charge-variance}
\end{equation}
so the number entropy is controlled by the growth (or boundedness) of charge fluctuations under the
drive.

\begin{figure}[t]
  \centering
  \includegraphics[width=0.65\linewidth]{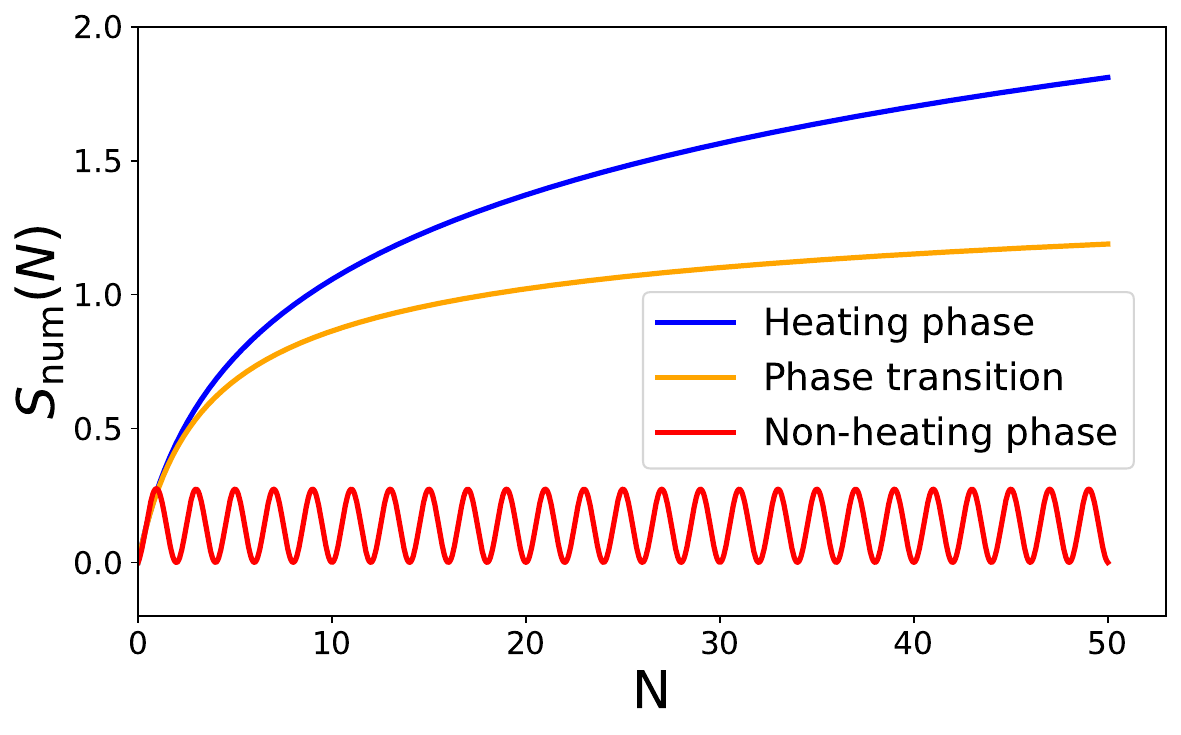}
  \caption{Evolution of the number (Shannon) entropy $S_{\rm num}(N)$ as a function of the number of drive cycles
  $N$ in the heating, non-heating, and phase-transition regimes of the two-step driving protocol~\eqref{eq:drive-two-step-drive-protocol}. Here we fix $k=3$ and $\theta=0.1$.}
  \label{fig:sree-number-entropy-phases}
\end{figure}

Using \eqref{eq:sree-number-entropy-definition} and \eqref{eq:sree-fixed-charge-moment-gaussian-approx} and replacing the sum by
an integral, one finds
\begin{align}
S_{\rm num}(N) &\simeq -\int_{-\infty}^{\infty} dq\; \mathcal Z_1(q;N)\,\log \mathcal Z_1(q;N) \nonumber
\\[2pt] &\simeq\; \log\!\big(2\sqrt{\pi B_1(N)}\big)+\frac12 . \label{eq:sree-number-entropy-gaussian-result}
\end{align}
The last expression highlights that (within this approximation) $S_{\rm num}$ is essentially the
differential entropy of a Gaussian probability distribution with variance $2B_1(N)$. Consequently, the qualitative distinct behaviour of \(S_{\rm num}(N)\) across the three Floquet regimes follows directly from the stroboscopic scaling of the Gaussian width \(B_1(N)\) (equivalently, of \(\log\mathcal D_N\) through \eqref{eq:sree-Znalpha-Bn}). The stroboscopic evolution of the number entropy with the drive cycles is shown in Fig.~\ref{fig:sree-number-entropy-phases}, and summarized in Table~\ref{tab:sree-Snum-scaling-phases}.
\begin{table}[t]
\centering
\renewcommand{\arraystretch}{1.2}
\setlength{\tabcolsep}{8pt}
\begin{tabular}{|c|c|c|}
\hline
\textbf{Driven CFT Phases} & \textbf{Width \boldmath \(B_1(N)\)} & \textbf{Number entropy \boldmath \(S_{\rm num}(N)\)} \\ 
\hline
Heating & \(B_1(N)\propto N\) & \(S_{\rm num}(N)\sim \log N\) \\ 
\hline
Phase transition & \(B_1(N)\propto \log N\) & \(S_{\rm num}(N)\sim \log(\log N)\) \\ 
\hline
Non-heating & \(B_1(N)\) $\sim$ \emph{oscillatory} & \(S_{\rm num}(N)\) $\sim$ \emph{bounded / oscillatory} \\ 
\hline
\end{tabular}
\caption{Asymptotic scaling of the Gaussian width \(B_1(N)\) (charge variance) and the corresponding number (Shannon) entropy \(S_{\rm num}(N)\) across the three drive regimes.}
\label{tab:sree-Snum-scaling-phases}
\end{table}

It is also useful to connect the number entropy to the difference between the total entanglement
entropy \(S_A(N)\) and the fixed-charge SREE. Using the Gaussian fixed-charge expression~\eqref{eq:sree-vn-fixed-charge}, one finds using \eqref{eq:sree-number-entropy-gaussian-result},
\begin{equation}
\label{eq:sree-SA-minus-SREE}
S_A(N)-S_A(q,N)
=S_{\rm num}(N)-\mathcal{O}\!\left(\frac{1}{B_1(N)}\right)
+\mathcal{O}\!\left(\frac{q^2}{B_1(N)}\right)\,.
\end{equation}
Thus, in the regime \(B_1(N)\gg 1\), the leading separation between the total entropy and the
fixed-charge SREE is set by the Shannon (number) entropy, while the remaining terms include an
explicit \(q\)-dependence that is parametrically suppressed by inverse powers of \(B_1(N)\).

\section{Non-unitary Quench}
\label{sec:nonunitary_quench}

In this section,  we explore the symmetry-resolved entanglement entropy under a general
non-unitary time evolution. We will primarily focus on the examples that have recently been considered
in Refs.~\cite{wen2024exactly, Bai:2026avl, Mao:2025cfl, Lapierre:2025zsg}, however, we will also offer some simple generalizations.

To this end, instead of applying the twist field formalism of the previous section, we will consider the free compact boson theory on a strip.\footnote{For the sake of completeness, we will review some elementary aspects, keeping the details at a bare minimum.} This is standard in the framework of global or local quenches, in which, the boundary of the strip encodes a non-trivial boundary state which is then evolved with the CFT Hamiltonian. This framework allows us to pose precise questions regarding the physics of the dynamical evolution, by analyzing how various correlation functions behave~\cite{Calabrese_2006, calabrese07quenches, calabrese07local, Calabrese:2016xau}.

In particular, the reduced density matrix associated with a spatial line segment $A$ is calculated by evaluating the path integral on the corresponding Euclidean space-time. The $n$-th Rényi entropy is given by sewing $n$ copies of this path integral along the line segment. This yields an $n$-sheeted cover of the original Euclidean space-time with conical singularities at the entangling points. Topologically, this yields an annulus. As formalized in Ref.~\cite{Cardy:2016fqc}, for both local and global quenches, entanglement between two semi-infinite segments can also be computed by evaluating the annulus path integral for the CFT.

\subsection{Quantum Quench with Complex Metrics: The set-up}

Let us begin with an arbitrary initial state $\ket{\psi_0}$.
At $t=0$, the state is evolved by a complex-time 
(non-unitary) evolution generated by the CFT Hamiltonian:
\begin{equation}
\ket{\psi(t)} \;=\; e^{-i H_{\rm CFT}(1-i\mu)t}\,\ket{\psi_0},
\;
\qquad \mu>0 \ ,\label{eq:quench-complex-time-evolution}
\end{equation}
where $\mu$ is a real parameter. The possibility of this type of evolution follows from the KSW criterion for admissible complex space-time metrics, on which generic QFTs can be consistently defined~\cite{kontsevich21wick, Witten:2021nzp}. In particular, Eq.~\eqref{eq:quench-complex-time-evolution} corresponds to the complex metric
\begin{eqnarray}
    ds^2 = - \left( 1 \pm i \mu\right) dt^2 + dx^2 \ .\label{eq:quench-complex-metric}
\end{eqnarray}
The sign choice $\mu>0$ is fixed so that the KSW criterion is satisfied and the path integral on this metric is convergent~\cite{Witten:2021nzp}. For $\mu=0$, Eq.~\eqref{eq:quench-complex-time-evolution} reduces to the usual unitary real-time evolution, while for $\mu>0$ the
evolution is non-unitary due to the damping factor $e^{-\mu H_{\rm CFT}t}$. 
This non-unitary time evolution can be thought of as a realization of a post-selection measurement protocol, composed with a usual Lorentzian time-evolution. However, it is natural to also think of this as the Lorentzian evolution of a non-trivial state which is obtained from the Euclidean evolution of the seed state $\ket{\psi_0}$.

Here we initially prepare the CFT  in a short-range entangled state which can be described 
by a regularized conformal boundary state~\cite{Calabrese_2006, calabrese07quenches, Cardy:1986gw,Cardy:2004hm}:
\begin{equation}
  \ket{\psi_0}=e^{-\frac{\beta}{4}H_{\rm CFT}}|b\rrangle \ ,\qquad \beta>0 \ ,\label{eq:quench-regularized-boundary-state}
\end{equation}
and $|b\rrangle$ denotes a non-renormalizable  conformal boundary state that in our case must respect the ${\rm U}(1)$ symmetry. Notice that there is no reason to prepare the boundary state using a CFT Hamiltonian and evolve the state with a different quantization that corresponds to a different CFT Hamiltonian\footnote{For example, on the cylinder, the boundary state can be prepared with the standard elliptic-class Hamiltonian $H_{\rm CFT} \sim L_0 + {\bar L}_0$, which then can be evolved with a hyperbolic-class Hamiltonian: $H \sim L_1 + L_{-1} + {\rm h.c.}$.}. The time evolution of the symmetry-resolved entanglement entropy in the unitary case ($\mu=0$) has been studied in Ref.~\cite{parez21quasiparticle}.

For simplicity, let us take an infinite system, $L\to\infty$, and
consider as subsystem $A$ the half-line $A=[0, \infty)$. Following Ref.~\cite{wen2024exactly}, to
obtain the time evolution of $\rho_A(t)={\rm Tr}_B(\ket{\psi(t)}\bra{\psi (t)})$,
we introduce the state 
\begin{equation}\label{eq:state_tau12}
\ket{\psi(\tau_1, \tau_2)}=e^{-(\frac{\beta}{4}+\mu\tau_1+\tau_2)H_{\rm CFT}}|b\rrangle
\end{equation}
and eventually perform in the final step the analytic continuation 
\begin{equation}
\tau_1\mapsto t,\quad \tau_2\mapsto it\ .
\label{eq:anal_cont}
\end{equation}

\begin{figure}[t]
 \includegraphics[width=\textwidth]{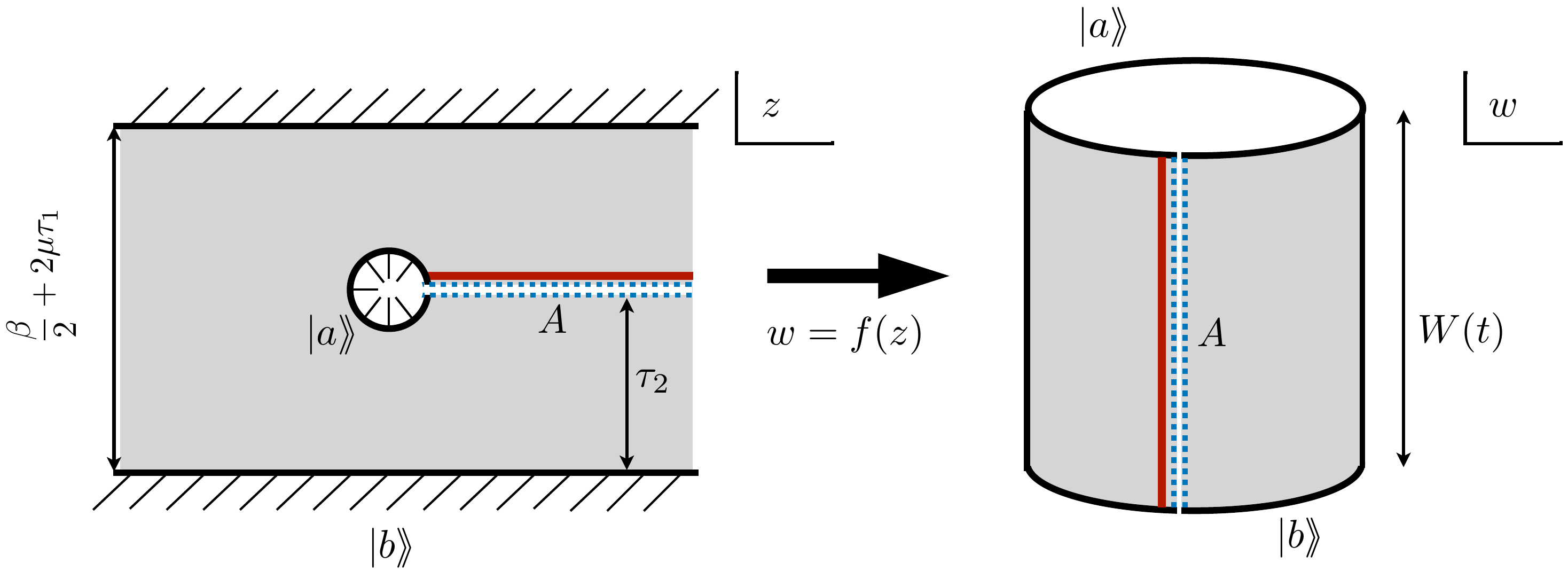}
\caption{Euclidean path-integral representation of the reduced density matrix of the state~\eqref{eq:state_tau12} for the half-line subsystem $A$ and its conformal mapping to the cylinder. The cylinder length $W(t)$ after the analytic continuation~\eqref{eq:anal_cont} is given by Eq.~\eqref{eq:quench-annulus-width-Wt}. The red line represents the line defect that implements the insertion $e^{i\alpha Q_A}$ in the charged moments $Z_n(\alpha; t)= \mathrm{Tr}\left(\rho_A(t)^{n}\,e^{i\alpha Q_A}\right) $.  }
 \label{fig:path_integral}
\end{figure}

The path integral representation of $\rho_A(\tau_1, \tau_2)={\rm Tr}_B(\ket{\psi(\tau_1, \tau_2)}\bra{\psi(\tau_1, \tau_2)})$ corresponds to 
the strip depicted in Fig.~\ref{fig:path_integral} left in the Euclidean space. By excising a small disk of radius $\epsilon_0$ at the 
end-point of $A$, as shown in the figure, we introduce a UV cutoff and a second boundary in the strip, which we assume 
is described by another conformally invariant boundary state $|a\rrangle$.  Using standard results (see e.g. Ref.~\cite{wen2024exactly, Cardy:2016fqc}), the strip geometry in Fig.~\ref{fig:path_integral} left can be conformally 
mapped to the cylinder in Fig.~\ref{fig:path_integral} right of circumference $2\pi$ and length $W(t)$ after the analytic continuation in Eq.~\eqref{eq:anal_cont}. Specifically, one finds\footnote{Given the complex plane to cylinder map, $w = f(z)$, where $w$ is the cylinder coordinate and $z$ is the complex plane coordinate, the cylinder length $W$ is given by $W = (1/2)({\cal W} + \bar{\cal W})$, where ${\cal W} = f(i\tau_2 + \infty) - f(i\tau_2 + \epsilon_0)$.}:
\begin{equation}
W(t)=\frac12\log\!\left[
\frac{\cosh\!\big(\frac{2\pi\epsilon_0}{\beta+4\mu t}\big)+
\cosh\!\big(\frac{4\pi t}{\beta+4\mu t}\big)}
{2\sinh^2\!\big(\frac{\pi\epsilon_0}{\beta+4\mu t}\big)}
\right] \ ,\qquad \mu\ge 0 \ .\label{eq:quench-annulus-width-Wt}
\end{equation}
Taking as explicit coordinates on the cylinder: 
\begin{equation}
 f(z)= w=u+iv,\qquad v\sim v+2\pi,\qquad u\in[0,W(t)]\,,\label{eq:quench-annulus-coordinates}
\end{equation}
the two ends of the cylinder at $u=0$ and $u=W(t)$ are endowed with the conformal boundary conditions encoded by the boundary states $|a\rrangle$ and $|b\rrangle$.

To obtain the entanglement entropy, we must compute ${\rm Tr}(\rho_A(t)^n)$. Following the previous steps, 
the path integral representation of ${\rm Tr}(\rho_A(t)^n)$ in Euclidean space corresponds to cyclically glue $n$ copies of the 
strip in Fig.~\ref{fig:path_integral} left along the subsystem $A$.  This replicated surface is conformally equivalent to a cylinder in Fig.~\ref{fig:path_integral} but with circumference $2\pi n$, 
and same length $W(t)$\footnote{It is easy to check that, after cyclically sewing together $n$
copies of the line segment corresponding to the subsystem $A$ on the $n$-sheeted cover of the Euclidean spacetime, the resulting geometry remains a punctured plane. Therefore, it is still topologically an annulus.}. Therefore, 
${\rm Tr}(\rho_A(t)^n)$ is identified with the partition function $\mathcal Z_n^{ab}(W)$ of the CFT on this cylinder with boundary conditions described by $|a\rrangle$ and $|b\rrangle$,
\begin{equation}
  \mathrm{Tr}(\rho_A^{\,n})=\frac{\mathcal Z_n^{ab}(W)}{\mathcal Z_1^{ab}(W)^n} \ .
  \label{eq:quench-renyi-trace-annulus}
\end{equation}
In the \emph{closed channel}, we quantize along the
$u$-direction; the evolution operator is, therefore, generated by the cylinder Hamiltonian (for circumference $2\pi$):
\begin{equation}
  H_{\rm cyl}=L_0+\bar L_0-\frac{c}{12} \ .
  \label{eq:quench-annulus-cylinder-hamiltonian}
\end{equation}
Thus, the cylinder partition function (of circumference $2\pi n$) is the closed-channel amplitude:
\begin{equation}
 \mathcal Z_n^{ab}(W)=\llangle a|\,e^{-H_{\rm cyl}W(t)/n}\,|b\rrangle \ .
  \label{eq:quench-annulus-partition-function-closed-channel}
\end{equation}
In the limit $W(t)\gg 1$, the Euclidean path integral is dominated by the vacuum and, therefore, yields the universal result~\cite{cc-04, Cardy:2016fqc}:
\begin{align}
  S_A^{(n)}(t) &=\frac{c}{12}\left(1+\frac1n\right)W(t)-(s_a+s_b) \ ,\label{eq:quench-renyi-entropy-annulus-W}
  \\
  S_A(t) &=\frac{c}{6}\,W(t)-(s_a+s_b) \ ,
  \label{eq:quench-von-neumann-entropy-annulus-W}
\end{align}
where the $W$-independent constants $s_{a(b)}=-\log(\llangle a (b)| 0\rangle)$, known as Affleck-Ludwig boundary entropies~\cite{PhysRevLett.67.161}, encode boundary/normalization data which are fixed by the
conformal boundary conditions.

At sufficiently long times:
\begin{equation}
W(t)\sim
\begin{cases}
\displaystyle \frac{2\pi}{\beta}\,t \ , & \mu=0 \ ,\\[6pt]
\displaystyle \log t \ , & \mu>0 \ \ (t\gg \beta/\mu) \ ,
\end{cases}\label{eq:quench-annulus-width-asymptotics}
\end{equation}
which implies that the R\'enyi entanglement entropies~\eqref{eq:quench-von-neumann-entropy-annulus-W} grow either linearly when $\mu=0$ (usual quench~\cite{calabrese05evolution}) or logarithmically in time when $\mu>0$ (non-unitary quench~\cite{wen2024exactly}).

Let us now comment on simple generalizations of the above non-unitary evolution. It is straightforward to check that the Witten-Kontsevich-Segal allowability condition also allows for a general class of metrics: 
\begin{eqnarray}
    ds^2 = - \left( 1 \pm i \mu(t)\right) dt^2 + dx^2 \ ,\label{eq:quench-complex-metric-eps-t}
\end{eqnarray}
where $\mu(t)$ is any function of $t$. This is a seemingly trivial statement, since the metric above is (complex-valued) diffeomorphic to the case of constant $\mu$. Nonetheless, the choice of the quantization is tied to the identification of a specific time coordinate and once this choice is made, a time reparametrization modifies the physics. This will finally result in a non-trivial time dependence of $W(t)$ in the regime $t \gg \beta / \mu(t)$. For example, suppose $\mu(t)$ is a non-decreasing function of $t$, then proceeding as above, we obtain:
\begin{eqnarray}
    W(t)\sim \log \left( \int dt\; \mu(t)  \right) \ , \quad {\rm with} \quad \int dt \; \mu(t) \gg \beta \ .\label{eq:quench-annulus-width-general-eps} 
\end{eqnarray}
One way to interpret the non-unitary evolution is to compose a Lorentzian time-evolution with a post-selection measurement, see {\it e.g.} Ref.~\cite{Bai:2026avl}. In particular, one considers the system coupled with an ancilla in an explicit and specific way, and starting with an appropriate tensor product state, a measurement is performed on the ancilla. Tuning the system and ancilla interaction appropriately corresponds to the Euclidean evolution of the non-unitary dynamics. The choice of the function $\mu(t)$ determines this interaction. For any positive and monotonically increasing function $\mu(t)$, an $e^{- H \int \mu(t)\; dt}$ operator receives a dominant contribution from the vacuum state in the large time limit. On the other hand, for a monotonically decreasing function $\mu(t)$, a similar vacuum dominance exists at early times. Therefore, the choice of the function $\mu(t)$, which is equivalent to choosing the quantization, can yield a rich set of possibilities. From the perspective of measurement-associated (or measurement induced) physics, this is an interesting aspect that we leave for a future work.

\subsection{SREE after a Quantum Quench with Complex Metrics}

Armed with the above framework, we will now explore the physics of symmetry resolved entanglement entropy under the non-unitary evolution in Eq.~\eqref{eq:quench-complex-time-evolution}.

As in previous sections, we consider the  free compact boson. To access symmetry-resolved quantities, we calculate first the charged moments $Z_n(\alpha; t)\equiv \mathrm{Tr}\!\left(\rho_A(t)^{n}\,e^{i\alpha Q_A}\right) $ and then project to fixed charge $q\in\mathbb Z$ by the Fourier transform in Eq.~\eqref{eq:sree-Znq-and-Znalpha-definitions}. The symmetry-resolved R\'enyi entropy is then given by Eq.~\eqref{eq:sree-renyi-via-Znq}. The path integral of $\rho_A(t)e^{i\alpha Q_A}$ is equivalent to that of $\rho_A(t)$, shown in Fig.~\ref{fig:path_integral}, but including a defect line along the cut of subsystem $A$ (which corresponds to the red line in the figure) that implements the insertion $e^{i\alpha Q_A}$. Therefore, the charged moments $Z_n(\alpha; t)$ are equivalent to the partition function of a cylinder of radius $2\pi n$ and length $W(t)$ with a defect line that connects the two boundaries of the cylinder:
\begin{equation}
Z_n(\alpha; t)=\frac{\mathcal{Z}_n^{ab}(W, \alpha)}{\mathcal{Z}_1^{ab}(W, 0)^n}.
\end{equation}
As noted in Refs.~\cite{DiGiulio:2022jjd,Kusuki:2023bsp}, the symmetry resolution of the entropy is only possible if both boundary states $|a\rrangle$, $|b\rrangle$ respect the ${\rm U}(1)$ symmetry.

In the closed string, this partition function can be written as the amplitude:
\begin{equation}
\mathcal{Z}_n^{ab}(\alpha; W)=\llangle a| e^{- H_{{\rm cyl}}(\alpha)W(t)/n}\,|b\rrangle \ ,
\label{eq:quench-annulus-closed-channel-fluxed-twist-amplitude}
\end{equation}
where $H_{\rm cyl}(\alpha)$ is the Hamiltonian of the compact boson on the cylinder with twist boundary conditions 
where the defect is inserted~\cite{DiGiulio:2022jjd,Kusuki:2023bsp}.  In the limit $W(t)\gg 1$, the fluxed-twist ground state $\ket{\alpha}$ dominates the amplitude
\eqref{eq:quench-annulus-closed-channel-fluxed-twist-amplitude}, yielding:
\begin{equation}\label{eq:part_func_twist_eg}
\mathcal{Z}_n^{ab}(\alpha; W)\simeq \mathfrak g_a^{(\alpha)}\,\mathfrak g_b^{(\alpha)}\, e^{- E_{\alpha}W(t)/n}\ ,
\end{equation}
where $\mathfrak g_{a\,(b)}^{(\alpha)}=\llangle a\,(b)|\alpha\rangle$ and $E_{\alpha}$ is the ground state energy of $H_{{\rm cyl}}(\alpha)$:
\begin{equation}
E_{\alpha}=\Delta_\alpha-\frac{1}{12}.
\end{equation}
Here $\Delta_\alpha$ is the conformal dimension of the vertex operator $V_\alpha$ associated with the 
line defect, \textit{i.e.} $\Delta_\alpha=\kappa\alpha^2/2$. Therefore:

\begin{equation}
 \mathcal Z_n^{ab}(\alpha;W)
  \simeq
  \mathfrak g_a^{(\alpha)}\,\mathfrak g_b^{(\alpha)}\,
  \exp\!\left(\frac{1}{12n}W(t)-\frac{\kappa}{2n}\alpha^2 W(t)\right) \ .
  \label{eq:quench-charged-moment-largeW-vacuum-dominance}
\end{equation}
 Since $\alpha$ plays the role of a background flat $U(1)$ connection (similar to a chemical potential source) localized on the subsystem cut, the boundary overlaps are expected to be smooth near
$\alpha=0$, provided the theory and the boundary conditions do not exhibit a singular response at vanishing holonomy. This results in analyticity around $\alpha=0$ and one can expand:
\begin{equation}
  \log\!\big[\mathfrak g_a^{(\alpha)}\mathfrak g_b^{(\alpha)}\big]
  =
  \log\!\big[\mathfrak g_a\mathfrak g_b\big]
  -\frac{1}{2}\Sigma_n\,\alpha^2+\mathcal O(\alpha^4) \ ,
  \qquad
  \Sigma_n\equiv
  -\partial_\alpha^2\log\!\big[\mathfrak g_a^{(\alpha)}\mathfrak g_b^{(\alpha)}\big]\Big|_{\alpha=0} \ .
 \label{eq:quench-boundary-overlap-small-flux-expansion}
\end{equation}
Keeping terms up to $\mathcal O(\alpha^2)$, the charged moments take the Gaussian form:
\begin{equation}
\mathcal  Z_n^{ab}(\alpha;W)=\mathcal Z_n^{ab}(0;W)\,e^{-B_n (t)\alpha^2}+\mathcal O(\alpha^4) \ ,
  \qquad
  B_n (t)=\frac{\kappa W(t)}{2n}+\frac{\Sigma_n}{2} \ .
  \label{eq:quench-charged-moments-gaussian-width-effective}
\end{equation}

It is clear from the discussion above that our subsequent results are perturbative in the flux $\alpha$. In particular, non-Gaussianities of the twisted Affleck-Ludwig boundary entropy  are not captured within what we describe next. The generic structure of the symmetry resolved entanglement measures will, therefore, be dominated by Gaussian integrals and, for large enough time, an equipartition is expected on general grounds.  Interestingly, even within the Gaussian approximation, equipartition of symmetry resolved entanglement is non-trivial. To this end, let us introduce the finite-window Gaussian Fourier transform:
\begin{equation}
  \mathcal I(B,q)\equiv
  \int_{-\pi}^{\pi}\frac{d\alpha}{2\pi}\,e^{-iq\alpha}\,e^{-B\alpha^2} \ ,
  \qquad (B>0,\ q\in\mathbb Z) \ ,
  \label{eq:quench-finite-window-gaussian-fourier-I}
\end{equation}
so that:
\begin{equation}
\mathcal  Z_n(q)=\mathcal Z_n^{ab}(0;W)\,\mathcal I\!\big(B_n (t),q\big) \ ,
  \qquad
\mathcal  Z_1(q)=\mathcal Z_1^{ab}(0;W)\,\mathcal I\!\big(B_1 (t),q\big) \ .
  \label{eq:quench-fixed-charge-moments-factorization-I}
\end{equation}
Using the definition of the symmetry-resolved  R\'enyi-$n$ entropy in Eq.~\eqref{eq:sree-renyi-via-Znq}, we obtain:
\begin{equation}
  S^{(n)}_{A,q}(t)
  = S_A^{(n)}(t)
  +\frac{1}{1-n}\Big[
    \log\mathcal I\!\big(B_n (t),q\big)-n\log\mathcal I\!\big(B_1 (t),q\big)
  \Big] \ .
  \label{eq:quench-renyi-decomposition-I}
\end{equation}

For a generic value of $W(t)$, computations are substantially involved. However, in the large $W(t)$ regime, they become much simpler. 

For large $B$, the error-function terms that appear in the closed form of $\mathcal I(B ,q)$ in \eqref{eq:quench-finite-window-gaussian-fourier-I}, approach unity up to corrections in $\mathcal O(e^{-\pi^2 B })$ and
\begin{equation}
\mathcal{I}(B , q)=\frac{e^{-q/(4B )}}{2\sqrt{\pi B }}+\mathcal{O}(e^{-\pi^2B })\ .
\end{equation}
Using the explicit form~\eqref{eq:quench-charged-moments-gaussian-width-effective} of $B_n(t)$, we then have:
\begin{multline}
\frac{1}{1-n}\Big[
\log \mathcal{I}\!\left(B_n(t),q\right)
- n \log \mathcal{I}\!\left(B_1(t),q\right)
\Big]
\;=\;
-\frac12 \log\!\big(2\pi \kappa W(t)\big)
-\frac{\log n}{2(n-1)}
\\
+\frac{n\big(\Sigma_1-\Sigma_n\big)}{2(1-n)\,\kappa W(t)}
+\frac{n\big(n\Sigma_n-\Sigma_1\big)}{2(1-n)\,[\kappa W(t)]^2}\,q^2
+\mathcal O\!\left(e^{-\pi^2 \kappa W(t)}\right) \ .
\label{eq:quench-I-term-largeW-estimate}
\end{multline}
The charge $q$ first appears only in terms of order $\mathcal O(W(t)^{-2})$ with further charge-dependent corrections arising at higher inverse powers of $W(t)$ 
as well as through terms that are exponentially suppressed in $W(t)$.

Combining \eqref{eq:quench-renyi-decomposition-I}--\eqref{eq:quench-I-term-largeW-estimate}, the leading large-$W$ behavior
of the symmetry-resolved R\'enyi entropies is:
\begin{multline}
S^{(n)}_{A,q}(t)
=
\frac{c}{12}\left(1+\frac1n\right)W(t)
-(s_a+s_b)
-\frac12 \log\!\big(2\pi \kappa W(t)\big)
-\frac{\log n}{2(n-1)}
\\
\quad
+\frac{n\big(\Sigma_1-\Sigma_n\big)}{2(1-n)\,\kappa W(t)}
+\frac{n\big(n\Sigma_n-\Sigma_1\big)}{2(1-n)\,[\kappa W(t)]^2}\,q^2
+\mathcal O\!\left(e^{-\pi^2\kappa W(t)}\right) \ .
\label{eq:quench-renyi-largeW-asymptotics}
\end{multline}
Taking the limit $n\to 1$ yields the symmetry-resolved entanglement entropy:
\begin{multline}
S_{A,q}(t)
\;=\;
\frac{c}{6}W(t)
-(s_a+s_b)
-\frac12 \log\!\big(2\pi \kappa W(t)\big)-\cfrac{1}{2}
\\
\quad
+\frac{\widetilde\Sigma_1}{2\kappa W(t)}
-\frac{\Sigma_1+ \widetilde\Sigma_1}{2[\kappa W(t)]^2}\,q^2
+\mathcal O\!\left(e^{-\pi^2 \kappa W(t)}\right) \ ,
\label{eq:quench-sree-largeW-asymptotics}
\end{multline}
where we have introduced the notation $
\widetilde\Sigma_1 \equiv \left.\frac{\partial \Sigma_n}{\partial n}\right|_{n=1}
$. 

Observe that this result is analogous to that of the driven system~\eqref{eq:sree-vn-fixed-charge}, replacing $B_1(N)$ by the cylinder length $W(t)$, \textit{cf.} Fig.~\ref{fig:path_integral}, which now controls the growth of both the total and the symmetry-resolved entropies. 
Recall that, by making use of the chain of conformal transformations that map the strip into the cylinder, $W(t)$ is given by Eq.~\eqref{eq:quench-annulus-width-Wt}.  

 As in the driven system, while for large $W$, equipartition follows, the departure from it is governed by the $W^{-2}$-term. Observe that this term originates from the Affleck-Ludwig boundary entropies for the compact boson Hamiltonian with twist boundary conditions~\eqref{eq:part_func_twist_eg}.  Both for the unitary global quench and for its non-unitary ($\mu>0$) deformation, $W(t)$ grows with time according to Eq.~\eqref{eq:quench-annulus-width-asymptotics}.
 This immediately implies charge equipartition in both cases at long times. 

 The result in Eq.~\eqref{eq:quench-sree-largeW-asymptotics} relies on approximating the cylinder partition function~\eqref{eq:quench-annulus-closed-channel-fluxed-twist-amplitude} by its ground-state contribution.
 For the free boson CFT, the full partition function can be computed, yielding an exact expression for the symmetry-resolved entropy as a function of $W(t)$~\cite{DiGiulio:2022jjd}, which contains terms linear and logarithmic in $W(t)$, as well as an infinite series of $e^{-W(t)}$. 
 


Finally, using the decomposition in Eq.~\eqref{eq:sree-entropy-decomposition-config-number} of the total entropy into configurational and number contributions, we can immediately identify the number entropy from Eq.~\eqref{eq:quench-sree-largeW-asymptotics} as
\begin{equation}
S_{\rm num}(t)=\frac{1}{2}\log(2\pi\kappa W(t))+\frac{1}{2}
\sim\left\{\begin{array}{ll} \log(t), & \mu=0,\\
\log\log(t),& \mu>0\,\, (t\gg \beta/\mu),\end{array}\right.
\end{equation}
which, together with Eq.~\eqref{eq:quench-annulus-width-asymptotics}, shows that it grows differently in non-unitary quenches compared to the unitary case~\cite{parez21quasiparticle}.

\section{Summary \& Discussion}
\label{sec:sum_discuss}

In this article, we studied the time evolution of symmetry-resolved entanglement entropies in two distinct settings: a bulk-driven CFT and a non-unitary quench, extending previous results on quantum quenches in Refs.~\cite{parez21quasiparticle, Parez:2021pgq, Chen:2025fzm}.
We focus on how symmetry-resolved Rényi entropies approach equipartition, or equivalently deviate from it. To demonstrate this behavior, we used the two-dimensional free compact boson CFT with a U(1) symmetry. 

The standard intuition behind the existence of an equipartition of symmetry resolved fine-grained entropies stems from the thermodynamic limit, in which the sub-system size is taken to be of the order of a macroscopic scale, while the UV cut-off remains hierarchically small. In this limit, the probability with which each charge sector contributes to the symmetry resolved quantities becomes uniform, with negligible differences that are suppressed by the inverse logarithmic-size of the sub-system. Clearly, the uniform distribution develops large charge-dependent features when the sub-system size becomes prohibitively small, {\it i.e.}~of the order of the UV cut-off. While this is an unphysical limit, we demonstrated that, by choosing to time-evolve the CFT with an appropriate Hamiltonian, an effective length-scale can be created such that, even in the thermodynamic limit, the charge-dependent deviations are only characterized by the effective length that can be much smaller compared to the sub-system size. As a result, we are now equipped with a free parameter that controls the departure from equipartition, even in the thermodynamic limit.

This additional parameter has a purely algebraic origin: Given the Virasoro algebra, it is the integer-valued labels of the generators of the algebra. We demonstrated, based on the free boson system but extending more generally, that upon choosing the CFT Hamiltonian as $H \sim L_k+L_{-k} + {\rm h.c.}$, this effective scale hierarchy can be induced. Furthermore, using an explicit mode-expansion for the free boson, or using the oscillator representation of the Virasoro algebra, we also demonstrated that the corresponding Hamiltonian can be interpreted in terms of modes that couple distant modes in the frequency domain. This class of Hamiltonians is of the Lüscher-Mack type \cite{Luscher:1974ez}, and the corresponding evolutions are characterized by thermal ({\it i.e.}~heating) physics. At the same time, the overall Floquet dynamics reveals a richer structure, since the restoration of charge equipartition depends on the phase of the drive: it is asymptotically restored in the heating phase and at the phase transition, whereas in the non-heating regime it is not generically restored.

At this point, several natural future directions emerge. It is a straightforward but interesting question to explore symmetry resolved entanglement properties in the context of a holographic system. This is especially rich in its ingredients, where a Chern-Simons formulation of the gravitational description relates holonomies around the Wilson lines as the flux phases due to the vertex operator insertion, see \cite{Zhao:2020qmn, Weisenberger:2021eby, Zhao:2022wnp,PhysRevLett.131.151601}. It will be interesting to explore how the Lüscher-Mack type Hamiltonian is realized within the Chern-Simons formulation, especially, how a corresponding intrinsic thermality is encoded here.\footnote{See {\it e.g.}~\cite{Das:2024lra} for a discussion on how the event horizon evolves in the holographic description.} It would be further interesting to see how a departure from the equipartition manifests itself in the formulation.

A related question can be asked for excited states in the theory. We have already considered the evolution of excited states which can be interpreted as a general non-unitary evolution of a seed state. The Euclidean evolution can be interpreted as preparing a non-trivial excited state starting from the seed state, while the details of the Euclidean evolution determine fine-grained data of this excited state. In $2d$ CFTs, however, a systematic study of excited states is perhaps naturally captured in terms of elements of the Verma module~\cite{capizzi20excited}. It will be interesting to understand how time evolution behaviour is classified across the modules.

To have a symmetry resolved notion, one must consider states and evolutions that have a conserved charge under the symmetry. All technical ingredients are currently available to carry out a systematic analysis of this question. This is especially interesting in the context of potential scar or squeezed states in the spectrum, see {\it e.g.}~\cite{Liska_2023} for discussions on similar states in CFT. Evidently, there will be states that break the symmetry spontaneously, in which case, it will be important to understand how entanglement asymmetry~\cite{ares23asymmetry} behaves, especially when the time-evolution is performed by the Lüscher-Mack type Hamiltonians. See, {\it e.g.}~\cite{Banerjee:2024zqb,Benini:2024xjv} for related works in this direction. 

It is also natural to consider non-Abelian symmetries. It is known~\cite{Calabrese_2021} that, {\it e.g.}~in the SU$(2)_k$ WZW model, equipartition among the different sectors labelled by the irreps of the algebra is also respected 
at the leading order. The equipartition is broken by a term independent of the subsystem size determined by the dimension of the irrep. It will be very interesting to understand better the detailed dynamics of a WZW model under the evolution of the Lüscher-Mack type Hamiltonians, especially so, since this is an integrable model and there is a tension between the thermalizing effect of the Lüscher-Mack Hamiltonian with the intrinsic dynamical data of the WZW model. We hope to address these aspects in the near future.

\section{Acknowledgements}

\noindent The authors would like to thank Diptarka Das, Sumit R.~Das, Krishnendu Sengupta for useful discussions and comments. J.D. would like to thank Saptaswa Ghosh for helpful discussions. J.D. would also like to thank the organizers of ISM 2025, held at the Indian Institute of Technology Bhubaneswar and NISER, where parts of this work were presented. A.K. acknowledges the support of the Humboldt Research Fellowship for Experienced Researchers by the Alexander von Humboldt Foundation and for the hospitality of Theoretical Physics III, Department of Physics and Astronomy, Julius-Maximilians-Universit\"{a}t W\"{u}rzburg and the support from the ICTP through the Associates Programme (2024-2030) during the course of this work. F.A. acknowledges support from the European Research Council under the Advanced Grant no. 101199196 (MOSE).


\bibliographystyle{JHEP}
\bibliography{biblio}

\providecommand{\href}[2]{#2}\begingroup\raggedright\begin{thebibliography}{100}

\bibitem{Holzhey94}
C.~Holzhey, F.~Larsen and F.~Wilczek, \emph{{Geometric and Renormalized Entropy in Conformal Field Theory}}, \href{https://doi.org/10.1016/0550-3213%2894%2990402-2}{\emph{Nucl. Phys. B} {\bfseries 424} (1994) 443}.

\bibitem{cc-04}
P.~Calabrese and J.~Cardy, \emph{Entanglement entropy and quantum field theory}, \href{https://doi.org/10.1088/1742-5468/2004/06/P06002}{\emph{J. Stat. Mech.} (2004) P06002}.

\bibitem{cc-09}
P.~Calabrese and J.~Cardy, \emph{Entanglement entropy and conformal field theory}, \href{https://doi.org/10.1088/1751-8113/42/50/504005}{\emph{J. Phys. A: Math. Theor.} {\bfseries 42} (2009) 504005}.

\bibitem{Caraglio08}
M.~Caraglio and F.~Gliozzi, \emph{Entanglement entropy and twist fields}, \href{https://doi.org/https://doi.org/10.1088/1126-6708/2008/11/076}{\emph{JHEP} {\bfseries 11} (2008) 076}.

\bibitem{Furukawa08}
S.~Furukawa, V.~Pasquier and J.~Shiraishi, \emph{{Mutual Information and Boson Radius in c = 1 Critical Systems in One Dimension}}, \href{https://doi.org/10.1103/PhysRevLett.102.170602}{\emph{Phys. Rev. Lett.} {\bfseries 102} (2009) 170602}.

\bibitem{calabrese09disjoint}
P.~Calabrese, J.~Cardy and E.~Tonni, \emph{Entanglement entropy of two disjoint intervals in conformal field theory}, \href{https://doi.org/https://doi.org/10.1088/1742-5468/2009/11/P11001}{\emph{J. Stat. Mech.} (2009) P11001}.

\bibitem{calabrese11disjoint}
P.~Calabrese, J.~Cardy and E.~Tonni, \emph{Entanglement entropy of two disjoint intervals in conformal field theory ii}, \href{https://doi.org/https://doi.org/10.1088/1742-5468/2011/01/P01021}{\emph{J.Stat.Mech.} (2011) P01021}.

\bibitem{calabrese05evolution}
P.~Calabrese and J.~Cardy, \emph{Evolution of entanglement entropy in one-dimensional systems}, \href{https://doi.org/https://iopscience.iop.org/article/10.1088/1742-5468/2005/04/P04010/meta}{\emph{J. Stat. Mech.} (2005) P04010}.

\bibitem{kaufman16thermalization}
A.M.~Kaufman, M.E.~Tai, A.~Lukin, M.~Rispoli, R.~Schittko, P.M.~Preiss et~al., \emph{Quantum thermalization through entanglement in an isolated many-body system}, \href{https://doi.org/https://doi.org/10.1126/science.aaf6725}{\emph{Science} {\bfseries 353} (2016) 794}.

\bibitem{alba17thermo}
V.~Alba and P.~Calabrese, \emph{Entanglement and thermodynamics after a quantum quench in integrable systems}, \href{https://doi.org/https://doi.org/10.1073/pnas.1703516114}{\emph{PNAS} {\bfseries 114} (2017) 7947}.

\bibitem{Calabrese_2006}
P.~Calabrese and J.~Cardy, \emph{Time dependence of correlation functions following a quantum quench}, \href{https://doi.org/10.1103/physrevlett.96.136801}{\emph{Phys. Rev. Lett.} {\bfseries 96} (2006) 136801}.

\bibitem{calabrese07quenches}
P.~Calabrese and J.~Cardy, \emph{Quantum quenches in extended systems}, \href{https://doi.org/10.1088/1742-5468/2007/06/P06008}{\emph{J. Stat. Mech.} (2007) P06008}.

\bibitem{calabrese07local}
P.~Calabrese and J.~Cardy, \emph{Entanglement and correlation functions following a local quench: a conformal field theory approach}, \href{https://doi.org/10.1088/1742-5468/2007/10/P10004}{\emph{J. Stat Mech.} (2007) P10004}.

\bibitem{Calabrese:2016xau}
P.~Calabrese and J.~Cardy, \emph{{Quantum quenches in 1 + 1 dimensional conformal field theories}}, \href{https://doi.org/10.1088/1742-5468/2016/06/064003}{\emph{J. Stat. Mech.} (2016) 064003}.

\bibitem{Oka2019-ep}
T.~Oka and S.~Kitamura, \emph{{Floquet Engineering of Quantum Materials}}, \href{https://doi.org/https://doi.org/10.1146/annurev-conmatphys-031218-013423}{\emph{Annu. Rev. Condens. Matter Phys.} {\bfseries 10} (2019) 387}.

\bibitem{Lindner:2011guf}
N.H.~Lindner, G.~Refael and V.~Galitski, \emph{{Floquet topological insulator in semiconductor quantum wells}}, \href{https://doi.org/10.1038/nphys1926}{\emph{Nature Phys.} {\bfseries 7} (2011) 490}.

\bibitem{PhysRevX.6.041001}
A.C.~Potter, T.~Morimoto and A.~Vishwanath, \emph{{Classification of Interacting Topological Floquet Phases in One Dimension}}, \href{https://doi.org/10.1103/PhysRevX.6.041001}{\emph{Phys. Rev. X} {\bfseries 6} (2016) 041001}.

\bibitem{NathanRudner2015}
F.~Nathan and M.S.~Rudner, \emph{{Topological singularities and the general classification of Floquet--Bloch systems}}, \href{https://doi.org/10.1088/1367-2630/17/12/125014}{\emph{New J. Phys.} {\bfseries 17} (2015) 125014}.

\bibitem{khemani2019briefhistorytimecrystals}
V.~Khemani, R.~Moessner and S.L.~Sondhi, \emph{A brief history of time crystals},  \href{https://arxiv.org/abs/1910.10745}{{\ttfamily 1910.10745}}.

\bibitem{Else_2020}
D.V.~Else, C.~Monroe, C.~Nayak and N.Y.~Yao, \emph{Discrete time crystals}, \href{https://doi.org/10.1146/annurev-conmatphys-031119-050658}{\emph{Annu. Rev. Condens. Matter Phys.} {\bfseries 11} (2020) 467}.

\bibitem{PhysRevLett.117.090402}
D.V.~Else, B.~Bauer and C.~Nayak, \emph{Floquet time crystals}, \href{https://doi.org/10.1103/PhysRevLett.117.090402}{\emph{Phys. Rev. Lett.} {\bfseries 117} (2016) 090402}.

\bibitem{Zaletel_2023}
M.P.~Zaletel, M.~Lukin, C.~Monroe, C.~Nayak, F.~Wilczek and N.Y.~Yao, \emph{Colloquium: Quantum and classical discrete time crystals}, \href{https://doi.org/10.1103/revmodphys.95.031001}{\emph{Rev. Mod. Phys} {\bfseries 95} (2023) 031001}.

\bibitem{PhysRevLett.118.030401}
N.Y.~Yao, A.C.~Potter, I.-D.~Potirniche and A.~Vishwanath, \emph{Discrete time crystals: Rigidity, criticality, and realizations}, \href{https://doi.org/10.1103/PhysRevLett.118.030401}{\emph{Phys. Rev. Lett.} {\bfseries 118} (2017) 030401}.

\bibitem{wen2018floquetconformalfieldtheory}
X.~Wen and J.-Q.~Wu, \emph{Floquet conformal field theory},  \href{https://arxiv.org/abs/1805.00031}{{\ttfamily 1805.00031}}.

\bibitem{Fan_2020}
R.~Fan, Y.~Gu, A.~Vishwanath and X.~Wen, \emph{{Emergent Spatial Structure and Entanglement Localization in Floquet Conformal Field Theory}}, \href{https://doi.org/10.1103/physrevx.10.031036}{\emph{Physical Review X} {\bfseries 10} (2020) 031036}.

\bibitem{PhysRevResearch.2.023085}
B.~Lapierre, K.~Choo, C.~Tauber, A.~Tiwari, T.~Neupert and R.~Chitra, \emph{{Emergent black hole dynamics in critical Floquet systems}}, \href{https://doi.org/10.1103/PhysRevResearch.2.023085}{\emph{Phys. Rev. Res.} {\bfseries 2} (2020) 023085}.

\bibitem{Han:2020kwp}
B.~Han and X.~Wen, \emph{{Classification of $SL_2$ deformed Floquet conformal field theories}}, \href{https://doi.org/10.1103/PhysRevB.102.205125}{\emph{Phys. Rev. B} {\bfseries 102} (2020) 205125} [\href{https://arxiv.org/abs/2008.01123}{{\ttfamily 2008.01123}}].

\bibitem{PhysRevB.103.224303}
B.~Lapierre and P.~Moosavi, \emph{{Geometric approach to inhomogeneous Floquet systems}}, \href{https://doi.org/10.1103/PhysRevB.103.224303}{\emph{Phys. Rev. B} {\bfseries 103} (2021) 224303}.

\bibitem{Wen:2020wee}
X.~Wen, R.~Fan, A.~Vishwanath and Y.~Gu, \emph{{Periodically, quasiperiodically, and randomly driven conformal field theories}}, \href{https://doi.org/10.1103/PhysRevResearch.3.023044}{\emph{Phys. Rev. Res.} {\bfseries 3} (2021) 023044} [\href{https://arxiv.org/abs/2006.10072}{{\ttfamily 2006.10072}}].

\bibitem{PhysRevResearch.2.033461}
B.~Lapierre, K.~Choo, A.~Tiwari, C.~Tauber, T.~Neupert and R.~Chitra, \emph{Fine structure of heating in a quasiperiodically driven critical quantum system}, \href{https://doi.org/10.1103/PhysRevResearch.2.033461}{\emph{Phys. Rev. Res.} {\bfseries 2} (2020) 033461}.

\bibitem{Choo:2022lgm}
K.~Choo, B.~Lapierre, C.~Kuhlenkamp, A.~Tiwari, T.~Neupert and R.~Chitra, \emph{{Thermal and dissipative effects on the heating transition in a driven critical system}}, \href{https://doi.org/10.21468/SciPostPhys.13.5.104}{\emph{SciPost Phys.} {\bfseries 13} (2022) 104} [\href{https://arxiv.org/abs/2205.02869}{{\ttfamily 2205.02869}}].

\bibitem{Fan_2021}
R.~Fan, Y.~Gu, A.~Vishwanath and X.~Wen, \emph{{Floquet conformal field theories with generally deformed Hamiltonians}}, \href{https://doi.org/10.21468/scipostphys.10.2.049}{\emph{SciPost Physics} {\bfseries 10} (2021) 049}.

\bibitem{Wen:2021mlv}
X.~Wen, Y.~Gu, A.~Vishwanath and R.~Fan, \emph{{Periodically, Quasi-periodically, and Randomly Driven Conformal Field Theories (II): Furstenberg's Theorem and Exceptions to Heating Phases}}, \href{https://doi.org/10.21468/SciPostPhys.13.4.082}{\emph{SciPost Phys.} {\bfseries 13} (2022) 082} [\href{https://arxiv.org/abs/2109.10923}{{\ttfamily 2109.10923}}].

\bibitem{Wen:2022pyj}
X.~Wen, R.~Fan and A.~Vishwanath, \emph{{Floquet's Refrigerator: Conformal Cooling in Driven Quantum Critical Systems}},  \href{https://arxiv.org/abs/2211.00040}{{\ttfamily 2211.00040}}.

\bibitem{Das:2022jrr}
S.~Das, B.~Ezhuthachan, A.~Kundu, S.~Porey, B.~Roy and K.~Sengupta, \emph{{Out-of-Time-Order correlators in driven conformal field theories}}, \href{https://doi.org/10.1007/JHEP08(2022)221}{\emph{JHEP} {\bfseries 08} (2022) 221} [\href{https://arxiv.org/abs/2202.12815}{{\ttfamily 2202.12815}}].

\bibitem{Das:2022pez}
S.~Das, B.~Ezhuthachan, A.~Kundu, S.~Porey, B.~Roy and K.~Sengupta, \emph{{Brane detectors of a dynamical phase transition in a driven CFT}}, \href{https://doi.org/10.21468/SciPostPhys.15.5.202}{\emph{SciPost Phys.} {\bfseries 15} (2023) 202} [\href{https://arxiv.org/abs/2212.04201}{{\ttfamily 2212.04201}}].

\bibitem{Das_2024}
D.~Das, S.R.~Das, A.~Kundu and K.~Sengupta, \emph{{Exactly solvable Floquet dynamics for conformal field theories in dimensions greater than two}}, \href{https://doi.org/10.1007/jhep09(2024)095}{\emph{JHEP} {\bfseries 09} (2024) 095}.

\bibitem{Lapierre25engineered}
B.~Lapierre, T.~Numasawa, T.~Neupert and S.~Ryu, \emph{Floquet engineered inhomogeneous quantum chaos in critical systems}, \href{https://doi.org/https://doi.org/10.1103/cn3z-vfgr}{\emph{Phys. Rev. B} {\bfseries 112} (2025) 104317}.

\bibitem{Das:2025wjo}
D.~Das, S.R.~Das, A.~Kundu and K.~Sengupta, \emph{{Dynamical phases of higher dimensional Floquet CFTs}}, \href{https://doi.org/10.21468/SciPostPhys.20.2.045}{\emph{SciPost Phys.} {\bfseries 20} (2026) 045} [\href{https://arxiv.org/abs/2504.00099}{{\ttfamily 2504.00099}}].

\bibitem{Erdmenger:2025chu}
J.~Erdmenger, J.~Kastikainen and T.~Schuhmann, \emph{{Driven inhomogeneous CFT as a theory in curved space-time}}, \href{https://doi.org/10.1007/JHEP02(2026)255}{\emph{JHEP} {\bfseries 02} (2026) 255} [\href{https://arxiv.org/abs/2508.18350}{{\ttfamily 2508.18350}}].

\bibitem{Fang:2025rie}
J.~Fang, Q.~Zhou and X.~Wen, \emph{{Phase transitions in quasiperiodically driven quantum critical systems: Analytical results}}, \href{https://doi.org/10.1103/PhysRevB.111.094304}{\emph{Phys. Rev. B} {\bfseries 111} (2025) 094304} [\href{https://arxiv.org/abs/2501.04795}{{\ttfamily 2501.04795}}].

\bibitem{mo25complex}
L.-H.~Mo, R.~Moessner and H.~Zhao, \emph{Complex and tunable heating in conformal field theories with structured drives via classical ergodicity breaking},  \href{https://arxiv.org/abs/2505.21712}{{\ttfamily 2505.21712}}.

\bibitem{wen2024exactly}
X.~Wen, \emph{{Exactly solvable non-unitary time evolution in quantum critical systems I: effect of complex spacetime metrics}}, \href{https://doi.org/10.1088/1742-5468/ad7c3d}{\emph{J. Stat. Mech.} (2024) 103103}.

\bibitem{kontsevich21wick}
M.~Kontsevich and G.~Segal, \emph{Wick rotation and the positivity of energy in quantum field theory}, \href{https://doi.org/https://doi.org/10.1093/qmath/haab027}{\emph{Q. J. Math.} {\bfseries 72} (2021) 673}.

\bibitem{Witten:2021nzp}
E.~Witten, \emph{{A Note On Complex Spacetime Metrics}},  \href{https://arxiv.org/abs/2111.06514}{{\ttfamily 2111.06514}}.

\bibitem{garratt23measurements}
S.J.~Garratt, Z.~Weinstein and E.~Altman, \emph{Measurements conspire nonlocally to restructure critical quantum states}, \href{https://doi.org/https://doi.org/10.1103/PhysRevX.13.021026}{\emph{Phys. Rev. X} {\bfseries 13} (2023) 021026}.

\bibitem{yang23measurements}
Z.~Yang, D.~Mao and C.-M.~Jian, \emph{Entanglement in a one-dimensional critical state after measurements}, \href{https://doi.org/https://doi.org/10.1103/PhysRevB.108.165120}{\emph{Phys. Rev. B} {\bfseries 108} (2023) 165120}.

\bibitem{murciano23measurements}
S.~Murciano, P.~Sala, Y.~Liu, R.S.K.~Mong and J.~Alicea, \emph{{Measurement-altered ising quantum criticality}}, \href{https://doi.org/https://doi.org/10.1103/PhysRevX.13.041042}{\emph{Phys. Rev. X} {\bfseries 13} (2023) 041042}.

\bibitem{weinstein23measurements}
Z.~Weinstein, R.~Sajith, E.~Altman,  and S.J.~Garratt, \emph{Nonlocality and entanglement in measured critical quantum ising chains}, \href{https://doi.org/https://doi.org/10.1103/PhysRevB.107.245132}{\emph{Phys. Rev. B} {\bfseries 107} (2023) 245132}.

\bibitem{sala24measurement}
P.~Sala, S.~Murciano, Y.~Liu and J.~Alicea, \emph{Quantum criticality under imperfect teleportation}, \href{https://doi.org/https://doi.org/10.1103/PRXQuantum.5.030307}{\emph{PRX Quantum} {\bfseries 5} (2024) 030307}.

\bibitem{patil24measurements}
R.A.~Patil and A.W.W.~Ludwig, \emph{{Highly complex novel critical behavior from the intrinsic randomness of quantum mechanical measurements on critical ground states—A controlled renormalization group analysis}},  \href{https://arxiv.org/abs/2409.02107}{{\ttfamily 2409.02107}}.

\bibitem{liu25measurements}
Y.~Liu, S.~Murciano, D.F.~Mross and J.~Alicea, \emph{Boundary transitions from a single round of measurements on gapless quantum states}, \href{https://doi.org/https://doi.org/10.1103/l4b7-h5cd}{\emph{Phys. Rev. Research} {\bfseries 7} (2025) 023293}.

\bibitem{su24dynamics}
L.~Su, A.~Clerk and I.~Martin, \emph{Dynamics and phases of nonunitary floquet transverse-field ising model}, \href{https://doi.org/https://doi.org/10.1103/PhysRevResearch.6.013131}{\emph{Phys. Rev. Res.} {\bfseries 6} (2024) 013131}.

\bibitem{Mao24Local}
W.~Mao, M.~Nozaki, K.~Tamaoka and M.T.~Tan, \emph{Local operator quench induced by two-dimensional inhomogeneous and homogeneous cft hamiltonians}, \href{https://doi.org/https://doi.org/10.1007/JHEP07%282024%29200}{\emph{JHEP} {\bfseries 07} (2024) 200}.

\bibitem{Lapierre:2025zsg}
B.~Lapierre, P.~Pelliconi, S.~Ryu and J.~Sonner, \emph{{Driven nonunitary dynamics of quantum critical systems}}, \href{https://doi.org/10.1103/lwrz-jxrr}{\emph{Phys. Rev. B} {\bfseries 112} (2025) 104322} [\href{https://arxiv.org/abs/2505.01508}{{\ttfamily 2505.01508}}].

\bibitem{Bai:2026avl}
C.~Bai, W.~Mao, M.~Nozaki, M.T.~Tan and X.~Wen, \emph{{Relaxation Process During Complex Time Evolution In Two-Dimensional Integrable and Chaotic CFTs}},  \href{https://arxiv.org/abs/2601.09290}{{\ttfamily 2601.09290}}.

\bibitem{laflorencie14spin}
N.~Laflorencie and S.~Rachel, \emph{Spin-resolved entanglement spectroscopy of critical spin chains and luttinger liquids}, \href{https://doi.org/10.1088/1742-5468/2014/11/P11013}{\emph{J. Stat. Mech.} (2014) P11013}.

\bibitem{PhysRevLett.120.200602}
M.~Goldstein and E.~Sela, \emph{Symmetry-resolved entanglement in many-body systems}, \href{https://doi.org/10.1103/PhysRevLett.120.200602}{\emph{Phys. Rev. Lett.} {\bfseries 120} (2018) 200602}.

\bibitem{xavier2018}
J.C.~Xavier, F.C.~Alcaraz and G.~Sierra, \emph{Equipartition of the entanglement entropy}, \href{https://doi.org/10.1103/PhysRevB.98.041106}{\emph{Phys. Rev. B} {\bfseries 98} (2018) 041106}.

\bibitem{lukin19exp}
A.~Lukin, M.~Rispoli, R.~Schittko, M.E.~Tai, A.M.~Kaufman, S.~Choi et~al., \emph{Probing entanglement in a many-body localized system}, \href{https://doi.org/https://doi.org/10.1126/science.aau0818}{\emph{Science} {\bfseries 364} (2019) 6437}.

\bibitem{bonsignori19}
R.~Bonsignori, P.~Ruggiero and P.~Calabrese, \emph{Symmetry resolved entanglement in free fermionic systems}, \href{https://doi.org/10.1088/1751-8121/ab4b77}{\emph{J. Phys. A: Math. Theor.} {\bfseries 52} (2019) 475302}.

\bibitem{Fraenkel19sym}
S.~Fraenkel and M.~Goldstein, \emph{Symmetry resolved entanglement: exact results in 1d and beyond}, \href{https://doi.org/10.1088/1742-5468/ab7753}{\emph{J. Stat. Mech.} (2020) 033106}.

\bibitem{Murciano:2019wdl}
S.~Murciano, G.~Di~Giulio and P.~Calabrese, \emph{{Symmetry resolved entanglement in gapped integrable systems: a corner transfer matrix approach}}, \href{https://doi.org/10.21468/SciPostPhys.8.3.046}{\emph{SciPost Phys.} {\bfseries 8} (2020) 046} [\href{https://arxiv.org/abs/1911.09588}{{\ttfamily 1911.09588}}].

\bibitem{Bonsignori:2020laa}
R.~Bonsignori and P.~Calabrese, \emph{{Boundary effects on symmetry resolved entanglement}}, \href{https://doi.org/10.1088/1751-8121/abcc3a}{\emph{J. Phys. A} {\bfseries 54} (2021) 015005} [\href{https://arxiv.org/abs/2009.08508}{{\ttfamily 2009.08508}}].

\bibitem{Murciano:2020vgh}
S.~Murciano, G.~Di~Giulio and P.~Calabrese, \emph{{Entanglement and symmetry resolution in two dimensional free quantum field theories}}, \href{https://doi.org/10.1007/JHEP08(2020)073}{\emph{JHEP} {\bfseries 08} (2020) 073} [\href{https://arxiv.org/abs/2006.09069}{{\ttfamily 2006.09069}}].

\bibitem{Horvath20boost}
D.X.~Horváth and P.~Calabrese, \emph{Symmetry resolved entanglement in integrable field theories via form factor bootstrap}, \href{https://doi.org/https://doi.org/10.1007/JHEP11(2020)131}{\emph{JHEP} {\bfseries 11} (2020) 131}.

\bibitem{Horvath21U1}
D.X.~Horváth, L.~Capizzi and P.~Calabrese, \emph{{U(1) symmetry resolved entanglement in free 1+1 dimensional field theories via form factor bootstrap}}, \href{https://doi.org/https://doi.org/10.1007/JHEP05(2021)197}{\emph{JHEP} {\bfseries 05} (2021) 197}.

\bibitem{Capizzi:2022jpx}
L.~Capizzi, O.A.~Castro-Alvaredo, C.~De~Fazio, M.~Mazzoni and L.~Santamar{\'\i}a-Sanz, \emph{{Symmetry resolved entanglement of excited states in quantum field theory. Part I. Free theories, twist fields and qubits}}, \href{https://doi.org/10.1007/JHEP12(2022)127}{\emph{JHEP} {\bfseries 12} (2022) 127} [\href{https://arxiv.org/abs/2203.12556}{{\ttfamily 2203.12556}}].

\bibitem{Castro-Alvaredo:2024azg}
O.A.~Castro-Alvaredo and L.~Santamar{\'\i}a-Sanz, \emph{{Symmetry-resolved measures in quantum field theory: A short review}}, \href{https://doi.org/10.1142/S0217984924300023}{\emph{Mod. Phys. Lett. B} {\bfseries 39} (2025) 2430002} [\href{https://arxiv.org/abs/2403.06652}{{\ttfamily 2403.06652}}].

\bibitem{capizzi20excited}
L.~Capizzi, P.~Ruggiero and P.~Calabrese, \emph{{Symmetry resolved entanglement entropy of excited states in a CFT}}, \href{https://doi.org/https://doi.org/10.1088/1742-5468/ab96b6}{\emph{J. Stat. Mech.} (2020) 073101}.

\bibitem{Hung21branes}
L.~Hung and G.~Wong, \emph{Entanglement branes and factorization in conformal field theory}, \href{https://doi.org/https://doi.org/10.1103/PhysRevD.104.026012}{\emph{Phys. Rev. D} {\bfseries 104} (2021) 026012}.

\bibitem{Calabrese_2021}
P.~Calabrese, J.~Dubail and S.~Murciano, \emph{{Symmetry-resolved entanglement entropy in Wess-Zumino-Witten models}}, \href{https://doi.org/10.1007/jhep10(2021)067}{\emph{JHEP} {\bfseries 10} (2021) 067}.

\bibitem{estienne21corrections}
B.~Estienne, Y.~Ikhlef and A.~Morin-Duchesne, \emph{Finite-size corrections in critical symmetry-resolved entanglement}, \href{https://doi.org/https://scipost.org/10.21468/SciPostPhys.10.3.054}{\emph{SciPost Phys.} {\bfseries 10} (2021) 054}.

\bibitem{Ares:2022gjb}
F.~Ares, P.~Calabrese, G.~Di~Giulio and S.~Murciano, \emph{{Multi-charged moments of two intervals in conformal field theory}}, \href{https://doi.org/10.1007/JHEP09(2022)051}{\emph{JHEP} {\bfseries 09} (2022) 051} [\href{https://arxiv.org/abs/2206.01534}{{\ttfamily 2206.01534}}].

\bibitem{Milekhin23charge}
A.~Milekhin and A.~Tajdini, \emph{{Charge fluctuation entropy of Hawking radiation: a replica-free way to find large entropy}}, \href{https://doi.org/https://doi.org/10.21468/SciPostPhys.14.6.172}{\emph{SciPost Phys.} {\bfseries 14} (2023) 172}.

\bibitem{Ghasemi23universal}
M.~Ghasemi, \emph{{Universal Thermal Corrections to Symmetry-Resolved Entanglement Entropy and Full Counting Statistics}}, \href{https://doi.org/https://doi.org/10.1007/JHEP05%282023%29209}{\emph{JHEP} {\bfseries 05} (2023) 209}.

\bibitem{DiGiulio:2022jjd}
G.~Di~Giulio, R.~Meyer, C.~Northe, H.~Scheppach and S.~Zhao, \emph{{On the boundary conformal field theory approach to symmetry-resolved entanglement}}, \href{https://doi.org/10.21468/SciPostPhysCore.6.3.049}{\emph{SciPost Phys. Core} {\bfseries 6} (2023) 049} [\href{https://arxiv.org/abs/2212.09767}{{\ttfamily 2212.09767}}].

\bibitem{Kusuki:2023bsp}
Y.~Kusuki, S.~Murciano, H.~Ooguri and S.~Pal, \emph{{Symmetry-resolved entanglement entropy, spectra {\&} boundary conformal field theory}}, \href{https://doi.org/10.1007/JHEP11(2023)216}{\emph{JHEP} {\bfseries 11} (2023) 216} [\href{https://arxiv.org/abs/2309.03287}{{\ttfamily 2309.03287}}].

\bibitem{foligno23torus}
A.~Foligno, S.~Murciano and P.~Calabrese, \emph{{Entanglement resolution of free Dirac fermions on a torus}}, \href{https://doi.org/https://doi.org/10.1007/JHEP03(2023)096}{\emph{JHEP} {\bfseries 03} (2023) 096}.

\bibitem{capizzi23defect}
L.~Capizzi, S.~Murciano and P.~Calabrese, \emph{Full counting statistics and symmetry resolved entanglement for free conformal theories with interface defects}, \href{https://doi.org/10.1088/1742-5468/ace3b8}{\emph{J. Stat. Mech.} (2023) 073102}.

\bibitem{PhysRevLett.131.151601}
C.~Northe, \emph{Entanglement resolution with respect to conformal symmetry}, \href{https://doi.org/10.1103/PhysRevLett.131.151601}{\emph{Phys. Rev. Lett.} {\bfseries 131} (2023) 151601}.

\bibitem{fossati23nonherm}
M.~Fossati, F.~Ares and P.~Calabrese, \emph{{Symmetry-resolved entanglement in critical non-Hermitian systems}}, \href{https://doi.org/https://doi.org/10.1103/PhysRevB.107.205153}{\emph{Phys. Rev. B} {\bfseries 107} (2023) 205153}.

\bibitem{Gaur:2023yru}
H.~Gaur and U.A.~Yajnik, \emph{{Multi-charged moments and symmetry-resolved R{\'e}nyi entropy of free compact boson for multiple disjoint intervals}}, \href{https://doi.org/10.1007/JHEP01(2024)042}{\emph{JHEP} {\bfseries 01} (2024) 042} [\href{https://arxiv.org/abs/2310.14186}{{\ttfamily 2310.14186}}].

\bibitem{Gaur:2024vdh}
H.~Gaur, \emph{{Total and symmetry resolved entanglement spectra in some fermionic CFTs from the BCFT approach}}, \href{https://doi.org/10.1007/JHEP09(2024)173}{\emph{JHEP} {\bfseries 09} (2024) 173} [\href{https://arxiv.org/abs/2402.07557}{{\ttfamily 2402.07557}}].

\bibitem{Northe:2025qcv}
C.~Northe, \emph{{Fermion parity resolution of entanglement}}, \href{https://doi.org/10.1007/JHEP12(2025)134}{\emph{JHEP} {\bfseries 12} (2025) 134} [\href{https://arxiv.org/abs/2509.03605}{{\ttfamily 2509.03605}}].

\bibitem{Bai:2025ysg}
C.~Bai, M.T.~Tan, B.~Lapierre and S.~Ryu, \emph{{Spatially structured entanglement from nonequilibrium thermal pure states}}, \href{https://doi.org/10.1103/sg51-1c1s}{\emph{Phys. Rev. B} {\bfseries 113} (2026) 064311} [\href{https://arxiv.org/abs/2510.25868}{{\ttfamily 2510.25868}}].

\bibitem{Mao:2025hkp}
W.~Mao and M.~Nozaki, \emph{{Time Ordering Effects and Destruction of Quasiparticles in Two-dimensional Holographic CFTs}},  \href{https://arxiv.org/abs/2508.07645}{{\ttfamily 2508.07645}}.

\bibitem{parez21quasiparticle}
G.~Parez, B.~R. and P.~Calabrese, \emph{Quasiparticle dynamics of symmetry resolved entanglement after a quench: the examples of conformal field theories and free fermions}, \href{https://doi.org/https://doi.org/10.1103/PhysRevB.103.L041104}{\emph{Phys. Rev. B} {\bfseries 103} (2021) L041104}.

\bibitem{Parez:2021pgq}
G.~Parez, R.~Bonsignori, R.~Bonsignori, P.~Calabrese and P.~Calabrese, \emph{{Exact quench dynamics of symmetry resolved entanglement in a free fermion chain}}, \href{https://doi.org/10.1088/1742-5468/ac21d7}{\emph{J. Stat. Mech.} (2021) 093102} [\href{https://arxiv.org/abs/2106.13115}{{\ttfamily 2106.13115}}].

\bibitem{Feldman:2019upn}
N.~Feldman and M.~Goldstein, \emph{{Dynamics of Charge-Resolved Entanglement after a Local Quench}}, \href{https://doi.org/10.1103/PhysRevB.100.235146}{\emph{Phys. Rev. B} {\bfseries 100} (2019) 235146} [\href{https://arxiv.org/abs/1905.10749}{{\ttfamily 1905.10749}}].

\bibitem{scopa22hydro}
S.~Scopa and D.X.~Horvath, \emph{Exact hydrodynamic description of symmetry-resolved rényi entropies after a quantum quench}, \href{https://doi.org/10.1088/1742-5468/ac85eb}{\emph{J. Stat. Mech.} (2022) 083104}.

\bibitem{Li25ballistic}
G.~Li, L.~Dupays and P.~Ruggiero, \emph{Charged moments and symmetry-resolved entanglement from ballistic fluctuation theory},  \href{https://arxiv.org/abs/2602.12185}{{\ttfamily 2602.12185}}.

\bibitem{Chen:2025fzm}
H.-H.~Chen, X.-L.~Zhou, J.~Yin and M.~Zhang, \emph{{Symmetry resolved entanglement entropy after an inhomogeneous quench}}, \href{https://doi.org/10.1007/JHEP06(2025)224}{\emph{JHEP} {\bfseries 06} (2025) 224} [\href{https://arxiv.org/abs/2504.14661}{{\ttfamily 2504.14661}}].

\bibitem{Zhao:2024ivy}
Z.-X.~Zhao, S.~He, H.~Ouyang, H.-a.~Zeng and Y.-X.~Zhang, \emph{{Relative R{\'e}nyi entropy under local quenches in 2D CFTs}}, \href{https://doi.org/10.1007/JHEP02(2026)195}{\emph{JHEP} {\bfseries 02} (2026) 195} [\href{https://arxiv.org/abs/2412.10735}{{\ttfamily 2412.10735}}].

\bibitem{Li:2024qpx}
H.-H.~Li and P.-Y.~Chang, \emph{{Phase transitions from heating to non-heating in $SU(1,1)$ quantum dynamics: Applications to Bose-Einstein condensates and periodically driven coupled oscillators}}, \href{https://doi.org/10.21468/SciPostPhysCore.8.1.018}{\emph{SciPost Phys. Core} {\bfseries 8} (2025) 018} [\href{https://arxiv.org/abs/2405.12558}{{\ttfamily 2405.12558}}].

\bibitem{Zamolodchikov:1986}
A.B.~Zamolodchikov, \emph{Two-dimensional conformal symmetry and critical four-spin correlation functions in the {Ashkin-Teller} model}, \href{https://doi.org/https://jetp.ras.ru/cgi-bin/e/index/e/63/5/p1061?a=list}{\emph{Sov. Phys. JETP} {\bfseries 63} (1986) 1061}.

\bibitem{Be_ken_2020}
M.~Beşken, S.~Datta and P.~Kraus, \emph{{Quantum thermalization and Virasoro symmetry}}, \href{https://doi.org/10.1088/1742-5468/ab900b}{\emph{J. Stat. Mech.} (2020) 063104}.

\bibitem{Besken:2019jyw}
M.~Be{\c{s}}ken, S.~Datta and P.~Kraus, \emph{{Semi-classical Virasoro blocks: proof of exponentiation}}, \href{https://doi.org/10.1007/JHEP01(2020)109}{\emph{JHEP} {\bfseries 01} (2020) 109}.

\bibitem{Jiang:2024hgt}
H.~Jiang and M.~Mezei, \emph{{New horizons for inhomogeneous quenches and Floquet CFT}}, \href{https://doi.org/10.1007/JHEP04(2025)025}{\emph{JHEP} {\bfseries 04} (2025) 025}.

\bibitem{Caputa:2022zsr}
P.~Caputa and D.~Ge, \emph{{Entanglement and geometry from subalgebras of the Virasoro algebra}}, \href{https://doi.org/10.1007/JHEP06(2023)159}{\emph{JHEP} {\bfseries 06} (2023) 159} [\href{https://arxiv.org/abs/2211.03630}{{\ttfamily 2211.03630}}].

\bibitem{Anand:2017dav}
N.~Anand, H.~Chen, A.L.~Fitzpatrick, J.~Kaplan and D.~Li, \emph{{An Exact Operator That Knows Its Location}}, \href{https://doi.org/10.1007/JHEP02(2018)012}{\emph{JHEP} {\bfseries 02} (2018) 012} [\href{https://arxiv.org/abs/1708.04246}{{\ttfamily 1708.04246}}].

\bibitem{Das:2024lra}
J.~Das and A.~Kundu, \emph{{Flowery horizons {\&} bulk observers: sl$^{(q)}$(2,{\,}{\ensuremath{\mathbb{R}}}), drive in 2d holographic CFT}}, \href{https://doi.org/10.1007/JHEP05(2025)035}{\emph{JHEP} {\bfseries 05} (2025) 035} [\href{https://arxiv.org/abs/2412.18536}{{\ttfamily 2412.18536}}].

\bibitem{Mao:2025cfl}
W.~Mao, A.~Miyata, M.~Nozaki and F.~Omidi, \emph{{Entanglement Dynamics by (Non-)Unitary Local Operator Quenches in a 2D Holographic CFT}},  \href{https://arxiv.org/abs/2512.18781}{{\ttfamily 2512.18781}}.

\bibitem{Cardy:2016fqc}
J.~Cardy and E.~Tonni, \emph{{Entanglement Hamiltonians in two-dimensional conformal field theory}}, \href{https://doi.org/10.1088/1742-5468/2016/12/123103}{\emph{J. Stat. Mech.} (2016) 123103}.

\bibitem{Cardy:1986gw}
J.L.~Cardy, \emph{{Effect of Boundary Conditions on the Operator Content of Two-Dimensional Conformally Invariant Theories}}, \href{https://doi.org/10.1016/0550-3213(86)90596-1}{\emph{Nucl. Phys. B} {\bfseries 275} (1986) 200}.

\bibitem{Cardy:2004hm}
J.L.~Cardy, \emph{{Boundary conformal field theory}},  \href{https://arxiv.org/abs/hep-th/0411189}{{\ttfamily hep-th/0411189}}.

\bibitem{PhysRevLett.67.161}
I.~Affleck and A.W.W.~Ludwig, \emph{Universal noninteger ``ground-state degeneracy'' in critical quantum systems}, \href{https://doi.org/10.1103/PhysRevLett.67.161}{\emph{Phys. Rev. Lett.} {\bfseries 67} (1991) 161}.

\bibitem{Luscher:1974ez}
M.~Luscher and G.~Mack, \emph{{Global Conformal Invariance in Quantum Field Theory}}, \href{https://doi.org/10.1007/BF01608988}{\emph{Commun. Math. Phys.} {\bfseries 41} (1975) 203}.

\bibitem{Zhao:2020qmn}
S.~Zhao, C.~Northe and R.~Meyer, \emph{{Symmetry-resolved entanglement in AdS$_{3}$/CFT$_{2}$ coupled to U(1) Chern-Simons theory}}, \href{https://doi.org/10.1007/JHEP07(2021)030}{\emph{JHEP} {\bfseries 07} (2021) 030} [\href{https://arxiv.org/abs/2012.11274}{{\ttfamily 2012.11274}}].

\bibitem{Weisenberger:2021eby}
K.~Weisenberger, S.~Zhao, C.~Northe and R.~Meyer, \emph{{Symmetry-resolved entanglement for excited states and two entangling intervals in AdS$_{3}$/CFT$_{2}$}}, \href{https://doi.org/10.1007/JHEP12(2021)104}{\emph{JHEP} {\bfseries 12} (2021) 104} [\href{https://arxiv.org/abs/2108.09210}{{\ttfamily 2108.09210}}].

\bibitem{Zhao:2022wnp}
S.~Zhao, C.~Northe, K.~Weisenberger and R.~Meyer, \emph{{Charged moments in W$_{3}$ higher spin holography}}, \href{https://doi.org/10.1007/JHEP05(2022)166}{\emph{JHEP} {\bfseries 05} (2022) 166} [\href{https://arxiv.org/abs/2202.11111}{{\ttfamily 2202.11111}}].

\bibitem{Liska_2023}
D.~Liska, V.~Gritsev, W.~Vleeshouwers and J.~Minář, \emph{Holographic quantum scars}, \href{https://doi.org/10.21468/scipostphys.15.3.106}{\emph{SciPost Physics} {\bfseries 15} (2023) 106}.

\bibitem{ares23asymmetry}
F.~Ares, S.~Murciano and P.~Calabrese, \emph{Entanglement asymmetry as a probe of symmetry breaking}, \href{https://doi.org/https://doi.org/10.1038/s41467-023-37747-8}{\emph{Nat. Comms.} {\bfseries 14} (2023) 2036}.

\bibitem{Banerjee:2024zqb}
T.~Banerjee, S.~Das and K.~Sengupta, \emph{{Entanglement asymmetry in periodically driven quantum systems}}, \href{https://doi.org/10.21468/SciPostPhys.19.2.051}{\emph{SciPost Phys.} {\bfseries 19} (2025) 051}.

\bibitem{Benini:2024xjv}
F.~Benini, V.~Godet and A.H.~Singh, \emph{{Entanglement asymmetry in conformal field theory and holography}}, \href{https://doi.org/10.1093/ptep/ptaf080}{\emph{Prog. Theor. Exp. Phys.} {\bfseries 2025} (2025) 6} [\href{https://arxiv.org/abs/2407.07969}{{\ttfamily 2407.07969}}].

\end{thebibliography}\endgroup
\end{document}